\documentclass[a4paper,11pt]{article}
\usepackage{etex}
\pdfoutput=1
\usepackage{jcappub}

\usepackage{amsmath,amssymb,graphicx}
\usepackage{soul}
\usepackage[latin1]{inputenc}
\usepackage{dsfont}

\usepackage{subfigure}

\usepackage{array}
\usepackage{mathtools}
\usepackage{cancel}

\usepackage{color}

\usepackage[numbers,sort&compress]{natbib}
\AtBeginDocument{}

\usepackage{multirow}

\usepackage{verbatim}

\usepackage{filecontents}

\usepackage{booktabs}

\newcommand{\be}{\begin{eqnarray}}
\newcommand{\en}{\end{eqnarray}}

\newcommand{\bepm}{\begingroup\setlength{\arraycolsep}{8pt} \renewcommand{\arraystretch}{1} \begin{pmatrix}}
\newcommand{\enpm}{\end{pmatrix} \endgroup}

\newcommand{\newc}{\newcommand}
\newc{\beq}{\begin{equation}}
\newc{\eeq}{\end{equation}}
\newc{\bea}{\begin{eqnarray}}
\newc{\eea}{\end{eqnarray}}
\newc{\simlt}{~\mbox{\smaller\(\lesssim\)}~}
\newc{\simgt}{~\mbox{\smaller\(\gtrsim\)}~}

\def\RK{R_{K^{\ast}}}
\def\RKK{R_{K^{(\ast)}}}

\def\KSS{K^{(\ast)}}

\usepackage[section, below, above]{placeins}

\usepackage{afterpage}

\begin{document}

\hyphenpenalty=7000
\tolerance=10000


\begin{titlepage}

\vspace*{-15mm}
\vspace*{0.7cm}

\begin{center}

{\Large {\bf
Non-universal $\mathbf{Z'}$ from $\mathrm{SO}(10)$ GUTs with vector-like family and the origin of neutrino masses
}
}\\[8mm]

Stefan Antusch$^{\star\dagger}$\footnote{Email: \texttt{stefan.antusch@unibas.ch}},
Christian Hohl$^{\star}$\footnote{Email: \texttt{ch.hohl@unibas.ch}},
Steve F.~King$^{\ddagger}$\footnote{Email: \texttt{s.f.king@soton.ac.uk}},
and Vasja Susi\v{c}$^{\star}$\footnote{Email: \texttt{vasja.susic@unibas.ch}}

\end{center}

\vspace*{0.20cm}

\centerline{$^{\star}$
\it Department of Physics, University of Basel,}
\centerline{\it
Klingelbergstr.\ 82, CH-4056 Basel, Switzerland}

\vspace*{0.4cm}

\centerline{$^{\dagger}$ \it
Max-Planck-Institut f\"ur Physik (Werner-Heisenberg-Institut),}
\centerline{\it
F\"ohringer Ring 6, D-80805 M\"unchen, Germany}

\vspace*{0.4cm}

\centerline{$^{\ddagger}$ \it
School of Physics and Astronomy, University of Southampton,}
 \centerline{\it SO17 1BJ Southampton, United Kingdom}

\vspace*{1.2cm}

\begin{abstract}
\noindent
A $Z'$ gauge boson with mass around the (few) TeV scale is a popular example of physics beyond the Standard Model (SM) and can be a fascinating remnant of a Grand Unified Theory (GUT). Recently, $Z'$ models with non-universal couplings to the SM fermions due to extra vector-like states have received attention as potential explanations of the present $R_K$, $\RK$ anomalies; this includes GUT model proposals based on the $\mathrm{SO}(10)$ group. In this paper we further develop GUT models with a flavour non-universal low scale $Z'$ and clarify several outstanding issues within them. First, we successfully incorporate a realistic neutrino sector (with linear and/or inverse low scale seesaw mechanism), which was so far a missing ingredient. Second, we investigate in detail their compatibility with the $R_K$, $\RK$ anomalies; we find that the anomalies do not have a consistent explanation within such models. Third, we demonstrate that these models have other compelling phenomenological features; we study the correlations between the flavour violating processes of $\mu\to 3e$ and $\mu$-$e$ conversion in a muonic atom, showing how a GUT imprint could manifest itself in experiments.

\end{abstract}
\vspace{1cm}
\end{titlepage}

\section{Introduction} \label{sec:introduction}

The presence of a low energy $Z'$ gauge boson arising from a spontaneously broken additional gauged $\mathrm{U}(1)'$
symmetry is a well explored possibility for physics beyond the Standard Model (BSM).
Such an extra gauged $\mathrm{U}(1)'$ could emerge from more unified gauge groups such as
the left-right symmetric model, Pati-Salam, $\mathrm{SO}(10)$ or
$E_6$~\cite{Langacker:2008yv}.
A recent comparison of $\mathrm{SO}(10)$ inspired $Z'$ models can be found in~\cite{King:2017cwv}.
Such models typically predict universal $Z'$ couplings to the three families of quarks and leptons,
thereby avoiding flavour changing neutral currents (FCNCs) mediated by tree-level $Z'$ exchange,
such as $K_0-\bar K_0$
mixing, $\mu$-$e$ conversion and $\mu\rightarrow 3e $ \cite{Langacker:2008yv}.

Recently, non-universal $Z'$ models have been invoked to account for
semi-leptonic $B$ decays which violate $\mu - e$ universality
\cite{Descotes-Genon:2013wba}.
The LHCb Collaboration results for $B\rightarrow \KSS l^+l^-$
report the $R_K$ \cite{Aaij:2014ora} and $\RK$ \cite{Bifani} ratios of $\mu^+ \mu^-$ to $e^+ e^-$ final states,
to be about $70\%$ of their expected values corresponding to a $4\sigma$ deviation from the Standard Model (SM).
Following the measurement of $\RK$ \cite{Bifani}, a number of phenomenological analyses of these data
\cite{Hiller:2017bzc}
favour an operator of the left-handed (L) form \cite{Glashow:2014iga,Hiller:2017bzc,DAmico:2017mtc}.
There is a large and growing body of literature on non-universal $Z'$ models as applied to explaining
these anomalies \cite{large}, with recent papers
following the $\RK$ measurement in~\cite{recent}.
An important challenge faced by these models is the requirement that they be anomaly free,
another challenge is to make a model consistent with Grand Unified Theory (GUT).
A simple way to achieve both goals is to add a complete vector-like fourth fermion family,
which is anomaly-free and
mixes with the ordinary three chiral anomaly-free families, leading to
flavour-changing $Z'$ interactions, with the idea suggested some years ago in \cite{Langacker:1988ur}
and recently developed in a model independent way and applied to the $\RKK$ anomalies in \cite{King:2017anf}.
A general formalism was developed and applied to
two concrete examples, including a supersymmetric
$\mathrm{SO}(10)$ Grand Unified Theory.
The idea is that a $\mathrm{U}(1)_X$ subgroup survives down to the $\mathrm{TeV}$ scale,
together with a vector-like family originating from incomplete
$10$ and $45$ dimensional representations of $\mathrm{SO}(10)$ \cite{King:2017anf}.
After mixing with the vector-like family, non-universal $Z'$ couplings are induced, which
could account for
the anomalous $B$ decay ratios
$R_K$ and $\RK$ \cite{King:2017anf}.
However the origin of neutrino mass and mixing
and the lepton flavour violating predictions were not discussed.
Independently of the motivation coming from the $\RKK$ anomalies, it is interesting to consider a non-universal $Z'$ of this kind.

In the present paper we further develop the class of $\mathrm{SO}(10)$ GUT models with a vector-like fourth family and induced non-universal $Z'$ couplings, and attempt to answer the open questions mentioned earlier. We focus on and resolve the following topics, listed here in order of importance:
\begin{enumerate}
\item 
	\textbf{Model building in the neutrino sector}: We construct the neutrino sector within such a class of models. We show that the constraints from $\mathrm{SO}(10)$ imply that a conventional seesaw mechanism is not possible. On the other hand, a linear seesaw mechanism~\cite{Malinsky:2005bi} and an inverse seesaw mechanism \cite{Dev:2009aw} can be successfully implemented, and done so consistently with a $\mathrm{TeV}$ scale gauged $\mathrm{U}(1)_X$, which arises from the embedding of $\mathrm{SU(5) \times {U}(1)_X}$ into $\mathrm{SO}(10)$. With the right-handed (sterile) neutrinos in reach of collider experiments, this can serve as an independent motivation for $\mathrm{SO}(10)$ models with low energy $Z'$ and vector-like family of fermions, as well as opening a window towards testing the seesaw mechanism in GUTs.
\item 
	\textbf{Inability to explain $\RKK$ anomalies}: In a general $Z'$ model with vector-like fermions, it is possible to explain the $\RKK$ anomaly with flavour non-universal couplings to the $Z'$. Implementation within $\mathrm{SO}(10)$ relates some of these couplings, making also predictions in other observables. Despite the $\RKK$ anomalies being the original motivation for the introduction of this class of models, we show that the explanation of $\RKK$ leads to other predictions inconsistent with observations. As an explanation of $\RKK$, the $\mathrm{SO}(10)$ models with low-energy $Z'$ and vector-like fermions are thus disfavored.   	
\item
	\textbf{Prospects for charged lepton flavour violating process searches}: The flavour-violating couplings of the $Z'$ to the fermions imply contributions to various flavour-violating processes. The underlying $\mathrm{SO}(10)$ symmetry and unification requirements further relate some of these processes, representing another interesting phenomenological feature of this class of models. We briefly discuss this point by studying the processes $\mu\to 3e$ and the $\mu$-$e$ conversion in a muonic atom; the model's implications on these processes can be tested in future experiments.
\end{enumerate}

The layout of the remainder of the paper is as follows: in Section~\ref{section:model} we define our $Z'$ model and discuss the mass matrices of light fermions, and in particular the neutrino mass matrix and the generation of neutrino masses, where we successfully embed into these models a low scale seesaw mechanism. Phenomenological topics are then found in Section~\ref{section:constraints}:
in Section~\ref{section:pheno-RKK}, we show the inability of this class of $\mathrm{SO}(10)$ models to explain the $\RKK$ anomalies. In Section~\ref{section:pheno-mue}, we investigate the decay $\mu\to 3e$ and $\mu$-$e$ conversion in a muonic atom, their rate depending on the $Z'$ mass, and how these two processes are related. Then we conclude. Finally, the paper also has Appendices~\ref{A} and \ref{Appendix:models}, dedicated to some of the technical model building details, which would interrupt the flow of the main body of the text if placed there.

\section{Non-universal $Z'$ and neutrino mass generation\label{section:model}}

\subsection{Definition of the model\label{section:model-general}}
We start the main body of the paper by defining the class of models under consideration. They are SUSY\footnote{Although our explicit constructions are performed within SUSY models for the sake of concreteness, some of the model building considerations can be applied to non-SUSY models as well.} $\mathrm{SO}(10)$ models, with a low energy $Z'$, vector-like family of fermions, and a realistic neutrino sector. First, though, we discuss again the background and motivation for the models we consider, a topic alluded to already in the Introduction.

One possible low-energy extension of the Standard Model gauge group is an additional $\mathrm{U}(1)'$ gauge group factor, which could get spontaneously broken at the scale of a few $\mathrm{TeV}$, thus implying the existence of a massive $Z'$ gauge boson at that scale. The phenomenology of these models becomes much richer if the existence of an additional 4th vector-like family is assumed. If the $\mathrm{U}(1)'$ quantum numbers of the 4th vector-like family are different than those of the first three families, $Z'$ couples non-universally to the three light mass eigenstates from the SM, as was pointed out in \cite{Langacker:1988ur}. 
Models with a low-energy $Z'$ and vector-like 4th family have recently attracted even more interest due to the apparent violation of lepton universality in the measurement of the $\RKK$ ratios. 

Various possibilities of non-universal $Z'$ models were considered in~\cite{King:2017anf} in the wake of $\RKK$ anomalous measurements. One challenge for model builders is to find natural UV completions of said $Z'$ models; one such possibility, for example, is to embed them in a GUT theory. In such a scenario, the various non-universal couplings in various fermion sectors are related, so one is rewarded with an increased predictivity. Since the extra $\mathrm{U}(1)'$ increases the rank of the gauge group to $5$, the group $\mathrm{SO}(10)$ lends itself as a natural unified group candidate.  
In fact, an $\mathrm{SO}(10)$ derived $Z'$ model was proposed in~\cite{King:2017anf}, but with no consideration of the neutrino sector. We remedy this situation in this paper by extending that model and investigating some of its properties; indeed, since
the representation $\mathbf{16}$ of $\mathrm{SO}(10)$ contains (in addition to all the SM fermions of one family) also a right-handed neutrino, any $\mathrm{SO}(10)$ GUT model should aspire to also be a theory of neutrinos.

In our $\mathrm{SO}(10)$ models, we make use of the following chain of embeddings of maximal subgroups:
\begin{align}
\mathrm{SO}(10)&\quad\supseteq\;\mathrm{SU}(5)\times\mathrm{U}(1)_X\quad\supseteq \;\underbrace{\mathrm{SU}(3)_C\times\mathrm{SU}(2)_L\times\mathrm{U}(1)_Y}_{\equiv G_{\text{SM}}}\times\mathrm{U}(1)_X.
\end{align}
We identify the $\mathrm{U}(1)_X$ factor as the additional $\mathrm{U}(1)'$ containing the $Z'$ gauge boson; unlike in a bottom-up $Z'$ model, this fixes all the $\mathrm{U}(1)_X$ charges of the particles once the choice of $\mathrm{SO}(10)$ representations is made. In order to provide an observable $Z'$ at energies reachable in the relatively near future, the $\mathrm{U}(1)_X$ is only broken at the scale $M_{Z'}$ of a few $\mathrm{TeV}$. Gauge symmetry is thus broken in two steps to the SM group $G_{\text{SM}}$:
\begin{align}
\mathrm{SO}(10)\xrightarrow{M_{\text{GUT}}}\;G_{\text{SM}}\times\mathrm{U}(1)_X\xrightarrow{M_{Z'}} \;G_{\text{SM}}.
\end{align}
A summary of our definition of the particle content of the models, which survives to the $M_Z'$ scale, is given in Table~\ref{tab:funfields}. The last column also includes the $\mathrm{SO}(10)$ origin of these fields, showing which GUT representations are essential for any such model to have: The fermion sector at the $\mathrm{SO}(10)$ level consists of three copies of $\mathbf{16}_{Fi}$, and additional representations $\mathbf{10}_F$, $\mathbf{45}_F$ and three copies of singlets $\mathbf{1}_{Fi}$. Also, the Higgs sector requires at least the representations $\mathbf{10}_H$, $\mathbf{16}_H$ and $\mathbf{\overline{16}}_H$. Other fields may be necessary to achieve the desired setup at the $Z'$ scale, and we consider some concrete examples later. 

For the purposes of any phenomenological analysis we thus remain agnostic about the exact model at the GUT scale. We instead consider an entire class of GUT-scale models, to which a model belongs if and only if its field content surviving to the scale $M_{Z'}$ is that of Table~\ref{tab:funfields}. We discuss below step-by-step the motivation for all the fields at the $\mathrm{SO}(10)$ level for the desired field content at $M_{Z'}$:
\begin{enumerate}
\item \textbf{Three chiral families}: $\mathbf{16}_{Fi}$\par
We use representations $\mathbf{16}_{Fi}$ (with family index $i=1,2,3$) of $\mathrm{SO}(10)$ to contain the three chiral families of the SM, as usual in $\mathrm{SO}(10)$ GUT. They decompose under \hbox{$\mathrm{SU}(5)\times\mathrm{U}(1)_X$} as
\begin{align}
	\mathbf{16}_{Fi} &\rightarrow (\mathbf{10},1)_i + (\overline{\mathbf{5}},-3)_i + (\mathbf{1},5)_i \;,
\end{align}
where the $\mathrm{U}(1)_X$ charges have to be normalized with an additional factor of $(2 \sqrt{10})^{-1}$. We also remind the reader at this point about two well-known $\mathrm{SU}(5)\to G_{\mathrm{SM}}$ decompositions, whose familiarity we will assume from now on:
	\begin{align}
	\mathbf{10}&\to(\mathbf{3},\mathbf{2},+\tfrac{1}{6})+(\mathbf{\overline{3}},\mathbf{1},-\tfrac{2}{3})+(\mathbf{1},\mathbf{1},+1),\\
	\mathbf{5}&\to (\mathbf{3},\mathbf{1},-\tfrac{1}{3})+(\mathbf{1},\mathbf{2},+\tfrac{1}{2}).
	\end{align}
\item \textbf{Vector-like family}: $\mathbf{10}_F+\mathbf{45}_F$ \par
Following \cite{King:2017anf}, the fourth vector-like family of fermions is embedded into the real representations $\mathbf{10}_F$ and $\mathbf{45}_F$ of $\mathrm{SO}(10)$, which decompose under $\mathrm{SU}(5)\times\mathrm{U}(1)_X$ as
\begin{align}
	\begin{split}
	\mathbf{10}_F &\rightarrow (\mathbf{5},-2) + (\overline{\mathbf{5}},2) \;, \\
	\mathbf{45}_F &\rightarrow  (\mathbf{10},-4) + (\overline{\mathbf{10}},4) + (\mathbf{1},0) + (\mathbf{24},0) \;.
	\end{split}
\end{align}
We assume that the $(\mathbf{1},0)$ and $(\mathbf{24},0)$ parts in $\mathbf{45}_F$ decouple (get large GUT scale masses) and that the fourth vector-like family with masses near the $\mathrm{TeV}$ scale consists of the following parts\footnote{Note that $\mathrm{SU}(5)$ is already broken at the $M_{\text{GUT}}$ scale. We use its representations only for the convenience of compact notation.} \cite{King:2017anf}:
\begin{align}
	(\mathbf{5},-2) + (\overline{\mathbf{5}},2) + (\mathbf{10},-4) + (\overline{\mathbf{10}},4) \;.
	\label{vectorfamily}
\end{align}
At the $\mathrm{SU}(5)$ level, the same $\mathbf{5}+\mathbf{\overline{5}}+\mathbf{10}+\mathbf{\overline{10}}$ vector-like family would have been obtained if it originated from an $\mathrm{SO}(10)$ pair $\mathbf{16}+\mathbf{\overline{16}}$. If constructing a non-universal $Z'$ model, however, the fourth family fermions need different $\mathrm{U}(1)_X$ charges than the first three families, requiring the fourth family to be in a different representation to the standard $\mathbf{16}$.

We emphasise that the single vector-like family in Eq.~\eqref{vectorfamily} includes quarks as well as leptons.\footnote{This may be compared to the $\mathrm{SO}(10)$ model in \cite{Hisano:2015pma}
where there are only three low energy $({\bf 5}, -2)+({\bf \overline{5}}, 2)$ representations mixing with the three chiral families
leading to flavour changing $Z'$ interactions.}
\item \textbf{Higgs sector}: $\mathbf{10}_H+\mathbf{16}_H+\mathbf{\overline{16}}_H$\par

The usual Higgs doublets $H_u$ and $H_d$ emerge dominantly from the representation $\mathbf{10}_H$, allowing for the usual Higgs Yukawa couplings. The triplets, which are part of the same $\mathrm{SO}(10)$ representations, need to be heavy due to proton decay constraints. We do not consider the details of SUSY breaking. We assume that supersymmetry is broken at some scale $M_{\text{SUSY}}$, which may be near the $M_{Z'}$ scale.

In order to break the $\mathrm{U}(1)_X$ and to obtain large mixing between the three chiral families and the fourth vector-like family, a pair $\mathbf{16}_H+\mathbf{\overline{16}}_H$ is needed. In each of the terms the $\mathrm{SU}(5)$-singlet component, which is charged under $\mathrm{U}(1)_X$, gets a vacuum expectation value (VEV) at the few $\mathrm{TeV}$ scale. 

We assume that the only surviving parts to scale $M_{Z'}$ of these Higgs representations are the $H_u$ and $H_d$ of the MSSM, which dominantly come from doublets in $\mathbf{10}_H$, as discussed later when considering DT splitting. 
\item \textbf{Three extra singlets for neutrino sector}: $\mathbf{1}_{Fi}$\par

The next consideration is how to make the model consistent with light neutrino masses. The implementation of a conventional type I seesaw mechanism~\cite{typeI} in a non-universal low-scale $Z'$ coming from a $\mathrm{SO(10)}$ GUT will necessarily fail for the reasons we now discuss. The Majorana masses for $\nu_i^c\nu_j^c$ would have to come from $\mathrm{SO}(10)$ invariant operators of the form $\mathbf{16}_F^{2}\cdot\mathbf{X}$, where $\mathbf{X}$ can stand for one of $3$ possible Higgs representations: $\mathbf{10}$, $\mathbf{120}$ or $\mathbf{\overline{126}}$. The first two options do not give a $\nu^{c}\nu^{c}$ term since they lack SM singlets, while the SM singlet in $\mathbf{\overline{126}}$ carries a $\mathrm{U}(1)_X$ charge. Under the assumption of a low scale $Z'$ at $\mathrm{TeV}$ scales, the $\mathrm{U}(1)_X$ breaking VEV of $\mathbf{\overline{126}}$ must be at that same scale as well. Furthermore, there is no possible suppression of Dirac type mass terms $\nu_{i}\nu^{c}_{j}$, since $\mathrm{SO}(10)$ symmetry directly relates them to those in the charged lepton mass matrix. Since the Dirac masses cannot be suppressed, a $\mathrm{TeV}$ seesaw scale is much too low to give correct masses for light neutrinos.

It is possible, however, to circumvent the above limitations. Since our model contains a $\mathbf{16}_H$ and $\overline{\mathbf{16}}_H$, we introduce three $\mathrm{SO}(10)$-singlets $\mathbf{1}_{Fi}$ in order get the three light neutrino masses from an inverse or linear seesaw mechanism \cite{Dev:2009aw,Malinsky:2005bi}. We shall show below that such a setup indeed works. 
\item \textbf{Matter parity}\par
An additional ingredient of the model is that we assume a $\mathbb{Z}_2$ matter parity. All representations in the fermionic sector (i.e.~the ones with index $F$: $\mathbf{16}_{Fi}$, $\mathbf{10}_F$, $\mathbf{45}_F$ and $\mathbf{1}_{Fi}$) are odd under matter parity, while all Higgs representations (i.e.~the ones with the index $H$: $\mathbf{10}_H$, $\mathbf{16}_H$, $\mathbf{\overline{16}}_H$) have even parity.

\end{enumerate}

We now specify how to prepare a setup with the above ingredients, so that the inverse and linear seesaw mechanisms, which make use of the $\mathbf{1}_F$ fermions, are not spoiled by the vector-like family. We also assume that the $\mathbf{24}+\mathbf{1}$ $\mathrm{SU}(5)$ parts in $\mathbf{45_F}$ are decoupled (we will discuss later how to achieve this by adding some additional terms). A minimal such version of a superpotential with the field content as enumerated above has the form
\begin{align}
\label{eq_superpotential_1}
	W &= W_H + W_\mathrm{Yuk} \;,\\
\label{eq_superpotential_2}
	\begin{split}
	W_\mathrm{Yuk} &= \epsilon\; \mathbf{1}_F^2 + m_{10}\; \mathbf{10}_F^2 + m_{45}\; \mathbf{45}_F^2 \\
	&\quad + y_1\; \mathbf{16}_F^2 \cdot \mathbf{10}_H \\
	&\quad + Y_1\; \mathbf{16}_F \cdot \mathbf{1}_F \cdot \overline{\mathbf{16}}_H + Y_2\; \mathbf{16}_F \cdot \mathbf{10}_F \cdot \mathbf{16}_H + Y_3\; \mathbf{16}_F \cdot \mathbf{45}_F \cdot \overline{\mathbf{16}}_H \;,
	\end{split}
\end{align}
where we remain agnostic about the explicit form of the Higgs sector and thus the form of $W_H$. In the Yukawa sector $W_\mathrm{Yuk}$ $\epsilon$, $m_{10}$ and $m_{45}$ are mass parameters and we label the dimensionless couplings by $y_i$ and $Y_i$ depending on whether the Higgs representation coupling to two fermions is $\{\mathbf{10}_H\}$ or $\{\mathbf{16}_H,\overline{\mathbf{16}}_H\}$. Note that for the sake of simplicity the family indices are neglected in Eq.~\eqref{eq_superpotential_2}. In fact $\epsilon$ and $y_1$ are $3\times3$ matrices and the $Y_i$ are vectors of length three. The fermion mass matrices, which arise due to such a Yukawa sector, will be studied in detail in later subsections.

\begin{table}[htb]
\centering
\footnotesize
\begin{tabular}{ c c c l l l}
\toprule
\multirow{2}{*}{\rule{0pt}{4ex}Field}	& \multicolumn{4}{c}{Representation}&\multirow{2}{*}{\rule{0pt}{-54ex}$\mathrm{SO}(10)$ origin} \\
\cmidrule(lr){2-5}
		& $\mathrm{SU}(3)_c$ & $\mathrm{SU}(2)_L$ & $\mathrm{U}(1)_Y$ &$\mathrm{U}(1)_X$&\\
\midrule
$Q_i$ 		 & $\mathbf{3}$ & $\mathbf{2}$ & $\phantom{-}1/6$ & $\phantom{-}1$&$\mathbf{16}_{Fi}$\\
$u^c_i$ 		 & $\mathbf{\overline{3}}$ & $\mathbf{1}$ & $-2/3$ & $\phantom{-}1$&$\mathbf{16}_{Fi}$\\
$d^c_i$ 		 & $\mathbf{\overline{3}}$ & $\mathbf{1}$ & $\phantom{-}1/3$ & $-3$&$\mathbf{16}_{Fi}$\\
$L_i$ 		 & $\mathbf{1}$ & $\mathbf{2}$ & $-1/2$ & $-3$&$\mathbf{16}_{Fi}$\\
$e^c_i$ 		 & $\mathbf{1}$ & $\mathbf{1}$ & $\phantom{-}1$ & $\phantom{-}1$&$\mathbf{16}_{Fi}$\\
$\nu^c_i$ 		 & $\mathbf{1}$ & $\mathbf{1}$ & $\phantom{-}1$ & $\phantom{-}5$&$\mathbf{16}_{Fi}$\\
\midrule
$H_u$ & $\mathbf{1}$ & $\mathbf{2}$ & $\phantom{-}1/2$ &$-2$&$\mathbf{10}_H$ (mostly)\\
$H_d$ & $\mathbf{1}$ & $\mathbf{2}$ & $-1/2$ &$\phantom{-}2$&$\mathbf{10}_H$ (mostly) \\
\midrule
\rule{0pt}{3ex}%
$Q_4$ 		 & $\mathbf{3}$ & $\mathbf{2}$ & $\phantom{-}1/6$ & $-4$&$\mathbf{45}_F$\\
$u^c_4$ 		 & $\mathbf{\overline{3}}$ & $\mathbf{1}$ & $-2/3$ & $-4$&$\mathbf{45}_F$\\
$d^c_4$ 		 & $\mathbf{\overline{3}}$ & $\mathbf{1}$ & $\phantom{-}1/3$ & $\phantom{-}2$&$\mathbf{10}_F$\\
$L_4$ 		 & $\mathbf{1}$ & $\mathbf{2}$ & $-1/2$ & $\phantom{-}2$&$\mathbf{10}_F$\\
$e^c_4$ 		 & $\mathbf{1}$ & $\mathbf{1}$ & $\phantom{-}1$ & $-4$&$\mathbf{45}_F$\\
\addlinespace[0.8\defaultaddspace]
\rule{0pt}{3ex}%
$\tilde{Q}_4$ 		 & $\mathbf{\overline{3}}$ & $\mathbf{\overline{2}}$ & $-1/6$ & $\phantom{-}4$&$\mathbf{45}_F$\\
$\tilde{u}^c_4$ 		 &  $\mathbf{3}$  & $\mathbf{1}$ & $\phantom{-}2/3$ & $\phantom{-}4$&$\mathbf{45}_F$\\
$\tilde{d}^c_4$ 		 &  $\mathbf{3}$  & $\mathbf{1}$ & $-1/3$ & $-2$&$\mathbf{10}_F$\\
$\tilde{L}_4$ 		 & $\mathbf{1}$ & $\mathbf{\overline{2}}$ & $\phantom{-}1/2$ & $-2$&$\mathbf{10}_F$\\
$\tilde{e}^c_4$ 		 & $\mathbf{1}$ & $\mathbf{1}$ & $-1$ & $\phantom{-}4$&$\mathbf{45}_F$\\
\midrule
$s_i$ & $\mathbf{1}$ & $\mathbf{1}$ & $\phantom{-}0$ &$\phantom{-}0$&$\mathbf{\phantom{1}1}_{Fi}$ \\
\bottomrule
\end{tabular}
\caption{The matter $(F)$ and Higgs $(H)$ sectors of the $\mathrm{SO}(10)$ model, with odd and even matter parity respectively, consist below the GUT scale of the usual
three chiral families of left-handed quarks and leptons $F_i$ ($i=1,2,3$)
from three $\mathbf{16}_{Fi}$ and two
Higgs doublets $H_{(u,d)}$ from  $\mathbf{10}_H$,
plus a fourth
vector-like family of left-handed fermions $F_4$, $\tilde{F}_4$ from the
real representations $\mathbf{10}_F$ and $\mathbf{45}_F$ of $\mathrm{SO}(10)$,
with $\mathrm{U}(1)_X$ charges as shown.
We also include three $\mathrm{SO}(10)$ singlets $s_i$ from
$\mathbf{1}_{F}$
in order to implement the linear or inverse seesaw mechanisms. Beside the gauge quantum numbers, fields also carry some global charge(s), forbidding certain interaction terms so that the superpotential is that of Eq.~\eqref{eq_superpotential_2}; we do not commit to a specific implementation of these charges, but provide three examples in Table~\ref{table:global-charges}.
}
\label{tab:funfields}
\end{table}

Assuming the $\mathrm{SO}(10)$ field content used in Eq.~\eqref{eq_superpotential_2}, there are certain additional superpotential terms that are in principle allowed by gauge invariance: 
\begin{align}
  y_2\; \mathbf{1}_F \cdot \mathbf{10}_F \cdot \mathbf{10}_H \qquad\text{and}\qquad y_3\; \mathbf{10}_F \cdot \mathbf{45}_F \cdot \mathbf{10}_H.
\end{align}
These $2$ operators are undesired\footnote{The two operator mix states in the neutrino mass matrix: neutrino-like fermions in the $\mathbf{10}_F$ are mixed with  $\mathbf{1}_F$ or the ones in $\mathbf{45}_F$. This prevents the neutrino mass mechanism to operate via inverse/linear seesaw; we solve potential problems arising from this by simply forbidding these couplings.} and can be forbidden by introducing a global $\mathrm{U}(1)$ or $\mathbb{Z}_n$ symmetry. Furthermore, the mechanism for decoupling the $\mathbf{24}+\mathbf{1}$ $\mathrm{SU}(5)$ parts in $\mathbf{45}_F$ from the vector-like states has also not been discussed. These considerations require an extension of the minimal field content and an assignment of all fields under a global symmetry. In our analysis in this paper, however, we would like to remain agnostic regarding the exact details of these mechanisms, since the way to achieve these is not unique, and we are only interested in the resulting theory at the $M_{Z'}$ scale.

In order to show that extensions which achieve the desired decoupling and forbid the unwanted terms actually exist, we do provide $3$ concrete realizations based on the discrete symmetry $\mathbb{Z}_4$; they are distinguished by the type of seesaw mechanism which operates (inverse seesaw only, linear seesaw only, or both). The field content of these models, as well as the $\mathbb{Z}_4$ charges of their $\mathrm{SO}(10)$ representations, are given in Table~\ref{table:global-charges}. Model I can be viewed as a realization of the general/default case, and this is the one we shall be discussing in the main body of the text; models II and III show that it is possible to isolate only one mechanism of neutrino mass generation, but their details are for the most part relegated to the Appendix. Each of these models introduces new fields and terms into the superpotential (see Eq.~\eqref{eq_superpotential_model_I}, \eqref{eq_superpotential_model_II} and \eqref{eq_superpotential_model_III} that are discussed later), in order to obtain the desired field content at the scale $M_{Z'}$.

To summarize, we realize GUT based models with a low energy $Z'$ and vector like-fermions as discussed above: we propose a class of SUSY $\mathrm{SO}(10)$ models, defined by their field content which survives to the $Z'$ scale as shown in Table~\ref{tab:funfields}, and Yukawa interactions for the surviving content coming from a superpotential in Eq.~\eqref{eq_superpotential_2}. A few explicit UV realizations of models in this broader class are listed in Table~\ref{table:global-charges}; they introduce extra representations and new interactions in the superpotential to obtain the desired field content at the $Z'$ scale; they also introduce a global symmetry to forbid unwanted terms in the superpotential, so as not to spoil the neutrino mass generation mechanism.

\begin{table}[htb]
\centering
\footnotesize
\begin{tabular}{cccc}
\toprule
\multirow{2}{*}{\rule{0pt}{2ex}$\mathrm{SO}(10)$ rep.}&I: general&II: linear only&III: inverse only\\
&$\mathbb{Z}_4$&$\mathbb{Z}_4$&$\mathbb{Z}_{4}$\\
\midrule
$\mathbf{16}_F$&$1$&$0$&$0$\\
$\mathbf{10}_F$&$0$&$0$&$0$\\
$\mathbf{45}_F$&$0$&$1$&$2$\\
$\phantom{\mathbf{1}}\mathbf{1}_F$&$0$&$1$&$2$\\
\addlinespace
$\mathbf{10}_H$&$2$&$0$&$0$\\
$\mathbf{16}_H$&$3$&$0$&$1$\\
$\mathbf{\overline{16}}_H$&$3$&$3$&$2$\\
\midrule
$\mathbf{210}_H$&$0$&$2$&$0$\\
$\mathbf{45}_H$&$2$&$1$&$1$\\
$\mathbf{54}_H$&$0$&$0$&$0$\\
$\mathbf{54}'_H$&/&$2$&/\\
\bottomrule
\end{tabular}
\caption{Examples of three models with global $\mathbb{Z}_4$ charge assignments for irreducible representations of $\mathrm{SO}(10)$. The fermion and Higgs fields necessary for the generation of neutrino masses from Eq.~\eqref{eq_neutrino_mass_matrix} and giving only the superpotential terms in Eq.~\eqref{eq_superpotential_2} are above the line, while the particular Higgs fields below the line are used in our examples to realize the doublet-triplet splitting, as well as the split in the $\mathbf{45}_F$, so that the $\mathbf{24}$ and $\mathbf{1}$ parts of $\mathrm{SU}(5)$ become heavy. We do not commit to any particular realization of these mechanisms, and therefore also not commit to having the representations below the line. In Example I, we have the general case when both the inverse and linear seesaw mechanisms are present; Examples II and III realize a linear seesaw only and inverse seesaw only scenarios, and are further discussed in Appendix~\ref{Appendix:models}.
}
\label{table:global-charges}
\end{table}

\subsection{The quark and charged lepton sectors\label{section:model-ude}}

Having defined the class of models in the previous subsection, we now discuss the form the Yukawa matrices take in the up, down and charged lepton sectors.  

We shall consider only the low energy states listed in Table~\ref{tab:funfields} of the effective theory surviving down to the $Z'$ scale. That consideration necessarily implies a mass splitting in the $\mathbf{45}_F$ representation between the fourth vector-like family and the states in the $(\mathbf{1},0)$ and $(\mathbf{24},0)$.
We provide $3$ example models where this splitting is achieved in Table~\ref{table:global-charges}, but do not commit ourselves to any one mechanism or model of the Higgs sector. In Example I, obtaining such a mass splitting involves introducing a $\mathbf{210}_H$ with even matter parity, which introduces new terms into the superpotential of Eq.~\eqref{eq_superpotential_2}. The appended Yukawa superpotential for model I, denoted by $W_{\mathrm{Yuk}}^\text{I}$, is now written as

\begin{align}
W_{\mathrm{Yuk}}^\text{I}&=W_{\mathrm{Yuk}}+Z_1\;\mathbf{45}_F^2 \cdot \mathbf{210}_H.\label{eq_superpotential_model_I}
\end{align}

Now taking $m_{45}$ in $W_{\mathrm{Yuk}}$ and the VEV of $\mathbf{210}_H$ at the GUT scale, and assuming the VEV is aligned in the $\mathrm{SU}(5)$-singlet direction, one can obtain the mass of the vector-like family $\mathbf{10}+\mathbf{\overline{10}}$ of $\mathrm{SU}(5)$ at the $\mathrm{TeV}$ scale by tuning the values of $\langle \mathbf{210}_H \rangle$ and $m_{45}$, while the $\mathbf{24}$ and $\mathbf{1}$ parts remain heavy. In addition to the $\mathbf{210}_H$, Model I also has representations
$\mathbf{45}_H$ and $\mathbf{54}_H$, which are not (significantly) involved in the Yukawa sector and are to be discussed later.

Example models II and III are similar, with III having a more restrictive pattern of mixing between SM and vector-like states. In the remainder of this section we continue to focus on Example I, whereas Examples II and III are further discussed in Appendices~\ref{Appendix:II} and \ref{Appendix:III}. Note that the main differences of interest between models I, II and III lie in the neutrino sector.

After electroweak (EW) symmetry breaking we obtain the mass matrices for the up- and down-type quarks $\mathbf{M}_u$ and $\mathbf{M}_d$, for the charged leptons $\mathbf{M}_e$ and also for the neutrinos $\mathbf{M}_\nu$. They include the three chiral families and the fourth vector-like family. Since the states coming from $(\mathbf{1},0)$ and $(\mathbf{24},0)$ in $\mathbf{45}_F$ have masses at the GUT scale, they can safely be ignored in the following considerations. Explicit results given below show that there is indeed a large mixing between the chiral families and the vector-like family in the left and right components in $\mathbf{M}_u$, $\mathbf{M}_d$ and $\mathbf{M}_e$. We provide their form when computed from Eq.~\eqref{eq_superpotential_2}:\footnote{We used the freedom to redefine the normalization of couplings and VEVs to absorb in them the Clebsch factors such as $2\sqrt{2}$, thus simplifying the expressions for mass matrices as much as possible. Once chosen, we stick to the same convention throughout the paper.}

\begin{align}
    \mathbf{M}_u&=
        \begin{pmatrix}
        y_{1} v_u & -Y_{3}\bar{v}_{L} & Y_{3} \bar{v}_{R} \\
        -Y_{3} \bar{v}_{L} & 0 & m_{45} \\
        Y_{3} \bar{v}_{R} & m_{45} & 0 \\
        \end{pmatrix},
    &\mathbf{M}_d&=
        \begin{pmatrix}
        y_{1} v_d & -Y_{2} v_{L} & Y_{3} \bar{v}_{R} \\
        0 & 0 & m_{45} \\
        Y_{2} v_{R} & m_{10} & 0 \\
        \end{pmatrix},
    &\mathbf{M}_e&=
        \begin{pmatrix}
        y_{1} v_d & 0 & Y_{2} v_{R} \\
        -Y_{2} v_{L} & 0 & m_{10} \\
        Y_{3} \bar{v}_{R} & m_{45} & 0 \\
        \end{pmatrix}.\label{eq:ude_mass_matrix}
\end{align}
\noindent

Above, we have defined $v_u$ and $v_d$ as the VEVs of the Higgs doublets $H_u$ and $H_d$, respectively, both (almost exclusively) from $\mathbf{10}_H$ and both at the electroweak scale. The VEVs of $(\mathbf{1},5)$ and $(\overline{\mathbf{5}},-3)$ of $\mathrm{SU}(5)\times\mathrm{U}(1)_X$ in $\mathbf{16}_H$ are labeled by $v_R$ and $v_L$, respectively. The $\bar{v}_R$ and $\bar{v}_L$ in $\overline{\mathbf{16}}_H$ are defined in an analogous way. The VEVs $v_R$ and $\bar{v}_R$ will be assumed to lie at the few TeV scale. $v_L$ and $\bar{v}_L$ are ``induced VEVs'' after EW symmetry breaking and are much smaller than the EW scale.\footnote{The size of the induced VEVs results from considerations of doublet-triplet splitting, see Appendix~\ref{A} for more details.}
The parameters $y_1$, $Y_2$, $Y_3$, $m_{10}$ and $m_{45}$ are defined in Eq.~\eqref{eq_superpotential_2}. Note that for the sake of readability, we wrote the matrices as $3\times 3$, while they are in fact $5\times 5$, since the $(1,1)$ block is actually $3\times 3$, corresponding to the usual $3$ family Yukawa matrix $y_1$. Note that Eq.~\eqref{eq_superpotential_2} implies the matrix $y_1$ is symmetric, eliminating the need to write transposes. Similarly, the parameters $Y_2$ and $Y_3$ are actually a $3\times 1$ row of parameters; we do not write transposes since their presence can be inferred from the dimension of the blocks where these parameters are present.

The matrices $\mathbf{M}_u$, $\mathbf{M}_d$ and $\mathbf{M}_e$ above are written in the basis
\begin{align}
W&=\mathbf{u}^T\,\mathbf{M}_u\,\mathbf{u^c}+\mathbf{d}^T\,\mathbf{M}_d\,\mathbf{d^c}+\mathbf{e}^T\,\mathbf{M}_e\,\mathbf{e^c}+\ldots,
\end{align}
\begin{align}
\mathbf{u}&=(u_i,u_4,\tilde{u}_4^c)^T,&\mathbf{d}&=(d_i,d_4,\tilde{d}_4^c)^T,&\mathbf{e}&=(e_i,e_4,\tilde{e}_4^c)^T,\nonumber\\
\mathbf{u^c}&=(u_j^c,u_4^c,\tilde{u}_4)^T,&\mathbf{d^c}&=(d_j^c,d_4^c,\tilde{d}_4)^T,&\mathbf{e^c}&=(e_j^c,e_4^c,\tilde{e}_4)^T.\label{eq:ude_states}
\end{align}

These are matrices in the superpotential $W$, whose singular value decomposition gives directly the matrices of physical masses for fermions. Concerning labels of particle states, $u_i$ and $d_i$ are in $Q_i$, $u_4$ and $d_4$ are in $Q_4$, $e_i$ is in $L_i$ and $e_4$ is in $L_4$. For the states of the columns, we have $\tilde{u}_4$ and $\tilde{d}_4$ in $\tilde{Q}_4$, while $\tilde{e}_4$ is in $\tilde{L}_4$.

Note that we have in each sector two vector-like pairs of representations, with one pair transforming as doublets and the other pair as singlets under $\mathrm{SU}(2)_L$. In the charged lepton sector, for example, we are adding both doublets $L_4\oplus \tilde{L}_4$, as well as singlets $e^{c}_4\oplus \tilde{e}^c_{4}$. Due to their quantum numbers under $\mathrm{SU}(3)_C\times\mathrm{U}(1)_{\text{EM}}$, the weak singlet state $\tilde{e}^{c}_4$ joins the row states $L_i$ and $L_4$, which are doublets, while the doublet $\tilde{L}_4$ joins the column states $e^{c}_i$ and $e^{c}_4$, which are $\mathrm{SU}(2)$ singlets. Due to this feature, the $(5,i)$ entry in the $5\times 5$ mass matrix $\mathbf{M}_e$ (where $i=1,2,3$ denotes the first three families) couples together two weak singlets, while the $(i,5)$ coupling couples two weak doublets. For that reason, confirmed by explicit calculation in Eq.~\eqref{eq:ude_mass_matrix}, both $(i,5)$ and $(5,i)$ entries have VEVs transforming as singlets under the SM group (but breaking $\mathrm{U}(1)_X$), i.e.~the VEVs $v_R$ and $\bar{v}_R$ located respectively in $\mathbf{16}_H$ and $\mathbf{\overline{16}}_H$. For analogous reasons, this feature of having a $5$th weak singlet state mixing with weak doublets in the basis for rows, and conversely a $5$th doublet state mixing with singlets in the basis for columns, is a feature in all mass matrices in Eq.~\eqref{eq:ude_mass_matrix}.

In our example models in Table~\ref{table:global-charges}, a $\mathbf{54}_H$ is introduced for doublet-triplet splitting (discussed in Appendix~\ref{A}). This representation would introduce a new operator $\mathbf{10}^{2}_F\cdot \mathbf{54}_H$ into the Yukawa sector superpotential in Eq.~\eqref{eq_superpotential_2}. We assume in these examples its coupling to be sufficiently small, so that it only introduces a small split between the $m_{10}$ parameters in the down-quark and lepton sectors, while still keeping them at $\mathrm{TeV}$ scale.

\subsection{The neutrino sector\label{section:model-neutrino}}
Let us now turn to the neutrino sector. We consider the basis
\begin{align}
	(\nu_{i},\,\nu^{c}_{i},\,s_i,\,\nu_{4},\,\tilde{\nu}_{4})\;,
\end{align}
where $\nu_{i}$ and $\nu^{c}_{i}$ are the states which correspond to the left and right-handed neutrinos in $\mathbf{16}_{Fi}$ and $s_i$ are the states associated with the $\mathbf{1}_{Fi}$. Furthermore, $\nu_{4}$ in $L_4$ labels the state of the left-handed neutrino in $(\mathbf{\bar{5}},2)$ of $\mathbf{10}_F$ and $\tilde{\nu}_{4}$ in $\tilde{L}_4$ is the corresponding state in $(\mathbf{5},-2)$ of $\mathbf{10}_F$. In this basis the neutrino mass matrix is given by\footnote{The matrix $\mathbf{M}_\nu$ is computed for the superpotential of Eq.~\eqref{eq_superpotential_2}. The full mass matrix should be  enlarged by extra states coming from the $\mathbf{1}$ and $\mathbf{24}$ of $\mathrm{SU}(5)$ located in $\mathbf{45}_F$. These states though are decoupled and live at the GUT scale, as already discussed for other sectors. The largest mixing between these decoupled states and the low energy states is a $\mathrm{TeV}$ level mixing coming from the operator $\mathbf{16}_F\cdot\mathbf{45}_F\cdot\mathbf{\overline{16}}_H$, which does not however change the qualitative picture of our discussion of neutrino mass generation. We thus omit the extra states from Eq.~\eqref{eq_neutrino_mass_matrix}.}

\begin{align}
\label{eq_neutrino_mass_matrix}
\mathbf{M}_\nu &=
\begin{pmatrix}
0 & y_1 v_u & Y_1 \bar{v}_L & 0 & Y_2 v_R \\
 y_1 v_u & 0 & Y_1 \bar{v}_R & 0 & Y_2 v_L \\
 Y_1 \bar{v}_L & Y_1 \bar{v}_R & \epsilon & 0 & 0 \\
 0 & 0 & 0 & 0 & m_{10} \\
 Y_2 v_R & Y_2 v_L & 0 & m_{10} & 0
\end{pmatrix} \;.
\end{align}

For the sake of readability we only show here one chiral family in the neutrino mass matrix. In fact the upper left $3\times3$ block
(which is all that is usually considered in the inverse or linear seesaw models)
has dimensions $9\times9$ if all three chiral families are included. The parameters $\epsilon$, $m_{10}$, $y_1$, $Y_1$ and $Y_2$ are defined as in Eq.~\eqref{eq_superpotential_2}. As discussed above, $m_{10}$, $v_R$ and $\bar{v}_R$ lie at the few $\mathrm{TeV}$ scale which allows for potentially large mixing of the chiral families with the vector-like family in the quark and charged lepton sectors and to break $\mathrm{U}(1)_X$.

In the limit $\epsilon,\bar{v}_L \rightarrow 0$ the neutrino mass matrix in Eq.~\eqref{eq_neutrino_mass_matrix} has one eigenvalue equal to zero and four of the order of a few $\mathrm{TeV}$, corresponding to the mass scale of the vector-like family. If a small value of $\bar{v}_L$ (i.e. $\bar{v}_L \ll v_u$) is switched on, the light left-handed neutrino masses are generated by the linear seesaw mechanism~\cite{Malinsky:2005bi}
\begin{align}
\label{eq_linear_seesaw}
	m_\nu & \sim y_1 \frac{v_u \bar{v}_L}{\bar{v}_R} \frac{m_{10}^2}{m_{10}^2 + Y_2^2 v_R^2} \sim y_1 \frac{v_u \bar{v}_L}{\bar{v}_R} \;.
\end{align}
In contrast, if a small non-zero value of $\epsilon$ (i.e.\ $\epsilon \ll v_u$) is switched on, the inverse seesaw mechanism~\cite{Dev:2009aw} applies:
\begin{align}
\label{eq_inverse_seesaw}
	m_\nu & \sim \epsilon \left( \frac{y_1 v_u}{Y_1 \bar{v}_R} \right)^2 \frac{m_{10}^2}{m_{10}^2 + Y_2^2 v_R^2} \sim \epsilon \left( \frac{y_1 v_u}{Y_1 \bar{v}_R} \right)^2 \;.
\end{align}
In general we expect both $\epsilon$ and $\bar{v}_L$ to be present, and neutrino mass generation gets simultaneous contributions from both the linear and inverse seesaw mechanisms. To get the correct scale of light neutrino masses of around $\mathcal{O}(0.1-1\,\mathrm{eV})$ from Eq.~\eqref{eq_linear_seesaw} and \eqref{eq_inverse_seesaw}, we need $\bar{v}_L$ to be around $\mathcal{O}(10\,\mathrm{eV})$, and $\epsilon$ to be around $\mathcal{O}(\mathrm{keV})$, with at least one of the two approaching this upper estimate. Curiously, the simplest mechanism for doublet triplet splitting in our models automatically induces the doublet VEV $\bar{v}_L$ to be of the correct scale for neutrino mass generation (see Appendix~\ref{A} for details), so the inverse seesaw contribution is not needed.

The same arguments on neutrino mass generation hold true if all three chiral families are considered and in this case we get three light states in the neutrino mass matrix. These results show that the usual upper left $3\times3$ block of the neutrino mass matrix in
Eq.~\eqref{eq_neutrino_mass_matrix}
($9\times9$ if all three chiral families are included)
plays a crucial role in determining neutrino mass and mixing.

Having successfully implemented the linear and inverse seesaw mechanisms consistent with light neutrino masses, we can pose the question whether the usual type I, II and III seesaw mechanisms play any role in our model. As already discussed earlier, a type I seesaw mechanism~\cite{typeI} would involve Majorana mass terms $\nu^{c}_i\nu^{c}_j$. The only operator of the form $\mathbf{16}_F^2\cdot\mathbf{X}$ in our model, as seen from Eq.~\eqref{eq_superpotential_2}, is that for $\mathbf{X}=\mathbf{10}_H$. Since $\mathbf{10}_H$ does not contain any SM singlet, there is no $\nu^{c}\nu^{c}$ term in our model, confirmed by the vanishing $(2,2)$ entry of $\mathbf{M}_\nu$ in Eq.~\eqref{eq_neutrino_mass_matrix}. A type II seesaw~\cite{typeII} requires a presence of scalars in SM representations $(1,3,\pm 1)$; such representations are present in $\mathbf{126}$ and $\mathbf{\overline{126}}$ representations of $\mathrm{SO}(10)$, but not in any of the Higgs representations of our models in Table~\ref{table:global-charges}. Finally, a type III seesaw mechanism~\cite{typeIII} requires fermions in the weak triplet SM representation $(1,3,0)$. The fermionic sector of our model has such fermions in the $\mathbf{45}_F$. The triplet is a part of the $\mathbf{24}$ of $\mathrm{SU}(5)$, which we successfully decoupled and lies at the GUT scale. The mass couplings of the triplet to the SM left-handed neutrinos $\nu_i$ are generated in our case, however, by the operator $\mathbf{16}_F\cdot\mathbf{45}_F\cdot\mathbf{\overline{16}}_H$; the VEV of the weak doublet in $\mathbf{\overline{16}}_H$ is only an induced one at scales below $\mathcal{O}(10\,\mathrm{eV})$, as already discussed. The type III contribution is thus completely negligible, and is in fact not visible in Eq.~\eqref{eq_neutrino_mass_matrix} since the triplet is part of omitted states. We have thus seen that all $3$ standard types of seesaw mechanism play no role in our model.

In Table~\ref{table:global-charges}, we provide examples of three concrete models demonstrating the discussed features of neutrino mass generation: Example I gives a model where the general case of both a linear seesaw and an inverse seesaw occurs: both $\bar{v}_L$ and $\epsilon$ are non-vanishing, and $\bar{v}_L$ is automatically of the correct scale (see Appendix~\ref{A}). In Example II, the mass term $\epsilon$ is forbidden by global charges, so that neutrino masses are generated by linear seesaw only. In Example III, the global charges forbid the doublet VEV $\bar{v}_L $ to be induced, so only the inverse seesaw applies. Examples II  and III are discussed in more detail in Appendices~\ref{Appendix:II} and \ref{Appendix:III}, respectively.

Based on the results of this section, we conclude that while regular (high-scale) seesaw cannot be implemented within an $\mathrm{SO}(10)$ based $Z'$ model with a vector-like family of fermions, low scale seesaw mechanisms, in particular inverse and/or linear seesaw, can be realized. Construction of the neutrino sector is the last missing ingredient for these models to provide a full description of all observed fermion sectors.

\section{Flavour phenomenology and related bounds\label{section:constraints}}

In this section we focus on the phenomenological aspects of the $\mathrm{SO}(10)$ based $Z'$ models with a vector-like family of fermions. In particular, we investigate the $\RKK$ anomalies in Section~\ref{section:pheno-RKK}, and $\mu\to 3e$ and $\mu$-$e$ conversion in Section~\ref{section:pheno-mue}. Both sections require though a prior discussion on the $5\times 5$ Yukawa matrices, which we now proceed with.

We take our mass matrices in the up, down and charged lepton sectors to be those of
Eq.~\eqref{eq:ude_mass_matrix}. It needs to be noted though that their exact form has certain potentially problematic relations, such as the $\mathrm{SU}(5)$ relation $\mathbf{M}_e=(\mathbf{M}_d)^T$. The particular form of the matrices and the relations between them hold though only at the GUT scale, and bad relations may turn out to be remedied once RGE running to the $Z'$ scale is taken into account. We can also envision the possibility that the Yukawa sector of the quark and charged lepton sectors is extended, e.g.~with an addition of representation $\mathbf{\overline{126}}$. A correction only in the $3\times 3$ block may be sufficient. Furthermore, the free parameters of such models could then be further restricted/reduced in number by a more elaborate theory of the Yukawa sector, e.g.~with discrete symmetries. All these possibilities clearly go beyond the scope of this paper, since any precise prediction would require selecting a particular model from within the class we consider, while our emphasis has actually been on the neutrino sector.

Despite the above limitation of the Yukawa sector, it turns out some general points of this class of models can still be extracted. In particular, this includes both the aforementioned inconsistency of the $\RKK$ anomalies, as well as some considerations for the processes of $\mu$-$e$ conversion  and $\mu\to 3e$ decays. For the latter case, we assume that the mass matrices of Eq.~\eqref{eq:ude_mass_matrix} get only small corrections, at least in the parts other than the upper-left $3\times 3$ SM block.

To connect our model with low energy processes, we need to know how the mass eigenstates of the Standard Model are related to the flavour eigenstates in Eq.~\eqref{eq:ude_states}. This information is contained in the mass matrices of Eq.~\eqref{eq:ude_mass_matrix}. We label the vector of flavour eigenstates by $\mathbf{q}'$, the vector of mass eigenstates by $\mathbf{q}$, and an intermediate basis by $\mathbf{q}''$, defined below. The diagonalization is then performed by plane rotations $\mathbf{R}_{ij}$ parameterized in the following way:
\begin{align}
\begin{split}
\mathbf{q}''&=(\mathbf{R}_{45}\mathbf{R}_{35}\mathbf{R}_{25}\mathbf{R}_{15})\;(\mathbf{R}_{34}\mathbf{R}_{24}\mathbf{R}_{14})\;\mathbf{q}',\\
\mathbf{q}&=(\mathbf{R}_{23}\mathbf{R}_{13}\mathbf{R}_{12})\;\mathbf{q}'',
\end{split}\label{eq:mixing-angle-parametrization}
\end{align}
where $\mathbf{R}_{ij}$ is a rotation in the plane of the $i$-th and $j$-th state, specified by the mixing angle $\theta_{ij}$ and given by the sub-matrix in the $i$-$j$ plane
\begin{align}
\mathbf{R}_{ij}:=\begin{pmatrix}\phantom{-}\cos\theta_{ij}&\sin\theta_{ij}\\ -\sin\theta_{ij}&\cos\theta_{ij}\end{pmatrix}.
\end{align}
In what amounts above to a parametrization of a $5\times 5$ unitary transformation, we have ignored the CP violating phases.

With the parametrization in Eq.~\eqref{eq:mixing-angle-parametrization}, we get mixing angles $\theta^{q}_{ij}$ in all sectors, since the vector $\mathbf{q}$ runs over $\mathbf{u}$, $\mathbf{d}$, $\mathbf{e}$, $\mathbf{u}^{c}$, $\mathbf{d}^{c}$ and $\mathbf{e}^{c}$, which corresponds to the left and right basis of the mass matrices of the up, down and charged lepton sectors. We fix the notation of the left angles to be $\theta_{ij}^{qL}$ and of the right angles to be $\theta_{ij}^{qR}$, where $q=u,d,e$.

The order of the rotation matrices in Eq.~\eqref{eq:mixing-angle-parametrization} is such that we first rotate to the $\mathbf{q}''$ basis by rotating away the heavy vector-like families in the $4$th and $5$th state of the matrix. After arriving to the basis $\mathbf{q}''$, we can either directly pass to the full mass eigenbasis $\mathbf{q}$ by performing the remaining rotations between the first three states, or use an effective field theory approach, where we integrate out the heavy $4$th and $5$th state with well defined masses at $M_{Z}'$, evolve with RGE to $M_{Z}$, and only then pass to the mass eigenbasis $\mathbf{q}$ by engaging the EW VEVs. Either way, the parametrization in Eq.~\eqref{eq:mixing-angle-parametrization} ensures that the angles $\theta_{12}$, $\theta_{13}$ and $\theta_{23}$ correspond to the convention used in the CKM mixing angles of the SM.

Notice that in Eq.~\eqref{eq:ude_mass_matrix} the following entries vanish or are very small (with induced doublet VEVs at the $\mathrm{eV}$ scale): $(4,4)$, $(5,5)$, $(i,4)$ and $(4,i)$, with $i=1,2,3$.  This form of the mass matrices implies that the $\theta_{i4}$ angles can remove the $(i,5)$ and $(5,i)$ entries, while $\theta_{45}$ then diagonalize the vector-like states. In our model, the parameters $m_{45}$, $m_{10}$, $v_R$ and $\bar{v}_R$ are roughly at the $M_{Z'}$ scale, so the $\theta_{i5}$ angles are needed only to remove the entries generated when rotating a vanishing entry with a non-zero $(i,j)$ entry in the $3\times 3$ block by one of the other angles. The $\theta_{i5}$ angles are thus rendered to be of the order $\sim y_1v_d/M_{Z'}$, which is negligibly small in part due to the small $y_1$ SM Yukawa couplings. We thus use the approximation 
\begin{align}
\theta_{15}\approx\theta_{25}\approx\theta_{35}\approx 0,\label{eq:theta5}
\end{align}
\noindent
while the $\theta_{i4}$ angles can be determined by looking at the ratio between the two non-zero entries in the $5$-th column and row, since the induced VEVs $v_L$ entries are negligible (see Appendix~\ref{A}):
\begin{align}
\tan\theta^{uL}_{i4}\sim \tan\theta^{uR}_{i4}\sim \tan\theta^{dL}_{i4}\sim \tan\theta^{eR}_{i4}&\sim (Y_3)_i \;\bar{v}_R/m_{45},\label{eq:theta4-1}\\
\tan\theta^{dR}_{i4}\sim \tan\theta^{eL}_{i4}&\sim (Y_2)_i \;v_R/m_{10}.\label{eq:theta4-2}
\end{align}

The reader is reminded that the parameters $Y_2$ and $Y_3$ are in fact $3\times 1$ rows, thus index $i$ refers to their specific entries. We see that the form of the matrices from Eq.~\eqref{eq:ude_mass_matrix}, even if taken as approximate, correlates some of the angles to be of at least the same scale.

The conventions given above, as well as the discussion regarding the negligible size of the $i$-$5$ angles and the inter-correlated size of the $i$-$4$ angles, will be used in the remainder of this section.

\subsection{The $\RKK$ anomalies and the bound from $B_s^0$-$\bar{B}^0_s$ mixing\label{section:pheno-RKK}}
The original motivation of the non-universal $Z'$ in~\cite{King:2017anf} was the explanation of the $\RKK$ anomalies \cite{Aaij:2014ora,Bifani}. We will show below that in the $\mathrm{SO}(10)$ model considered in this paper and in~\cite{King:2017anf}, the explanation of  the $\RKK$ anomalies is disfavored by other measurements.

We shall first consider the special case from reference~\cite{King:2017anf}, where the $\RKK$ anomalies should be most easily realized: we assume that only $\theta_{34}^{dL}$ mixing in the down-quark sector and $\theta_{14}^{eL}$ mixing in the lepton sector are present, while the couplings $Y_2$ and $Y_3$ in Eq.~\eqref{eq:ude_mass_matrix} are such that other $i4$ mixing angles are assumed zero: $\theta^{dL}_{14}=\theta^{dL}_{24}=\theta^{eL}_{24}=\theta^{eL}_{34}=0$. We shall later discuss the general case, and show that the same conclusions hold.

Among the interactions involving $Z'$ and fermions, we have the following terms involving the left-handed fields:
\begin{align}
\mathcal{L}_{Z'}&=C_{Z'bs}\;Z'_\kappa\,\bar{b}_L \gamma^{\kappa} s_L + C_{Z'\mu\mu}\;Z'_\kappa\bar{\mu}_L\gamma^\kappa\mu_L+C_{Z'ee}\;Z'_\kappa\bar{e}_L\gamma^\kappa e_L+\ldots,\label{eq:Zp-Lagrangian}
\end{align}
where the three couplings, following~\cite{King:2017anf}, are equal to
    \begin{align}
     C_{Z'bs}&=-\tfrac{5}{2\sqrt{10}}\;g'(s_{34}^{dL})^2(V'^\dagger_{dL})_{32},\label{eq:CZbs-coupling}\\
     C_{Z'\mu\mu}&=-\tfrac{3}{2\sqrt{10}}\;g',\\
     C_{Z'ee}&=-\tfrac{3}{2\sqrt{10}}\;g'(1-\tfrac{5}{3}(s_{14}^{eL})^2)\label{eq:CZee-coupling},
     \end{align}
where $g'$ is the coupling of the $Z'$ gauge boson (in the convention where the $X$ charges are normalized with the $(2\sqrt{10})^{-1}$ factor), $s_{34}^{dL}=\sin\theta^{dL}_{34}$, $s_{14}^{eL}=\sin\theta^{eL}_{14}$, where $\theta^{dL}_{34}$ and $\theta^{eL}_{14}$ are assumed to be the only non-zero mixing angles between the first 3 families and the vector-like 4th family in the quark and lepton sectors, respectively. The $(V'^\dagger_{dL})_{32}$ is the $32$ component of the matrix $V'^\dagger_{dL}$, which describes the transformation of the mass eigenstates to the flavour basis in the down sector after the vector-like states have been decoupled, i.e.
    \begin{align}
    b''_L&=(V'^\dagger_{dL})_{31}\,d_L+(V'^\dagger_{dL})_{32}\,s_L+(V'^\dagger_{dL})_{33}\,b_L,
    \end{align}
where $d_L,s_L,b_L$ are mass eigenstates, while $b''_L$ is the flavour state coupling to $Z'$ (defined after integrating out the heavy 4th and 5th states).

The $\RKK$ anomalies involve detecting a deficit in the ratio of branching ratios of the processes $B\to \KSS \mu^{+}\mu^{-}$ compared to $B\to \KSS e^{+} e^{-}$.
The new physics contribution is favored to come from left-handed fields~\cite{Glashow:2014iga,Hiller:2017bzc,DAmico:2017mtc}, whose couplings to $Z'$ we have written in Eq.~\eqref{eq:CZbs-coupling}--\eqref{eq:CZee-coupling}. The deficit in muons should be coming from a BSM operator mediating $b\to s\mu^{+}\mu^{-}$:
\begin{align}
\mathcal{L}_{\mathrm{BSM}}&=G_{b_L\mu_L}\,(\bar{b}_L\gamma^\kappa s_L)(\bar{\mu}_L\gamma_\kappa \mu_L),
\end{align}
with the anomaly measurement requiring $G_{b_L\mu_L}^{\mathrm{exp}}=-(31\,\mathrm{TeV})^{-2}$. This value is based on the fit of the BSM Wilson coefficient for the left-left operator to be $C^{\textrm{BSM}}_{b_L\mu_L}=-1.33$ and the relation $G_{b_L\mu_L}^{\mathrm{exp}}=C^{\textrm{BSM}}_{b_L\mu_L}/(36\,\mathrm{TeV})^2$, as described in~\cite{DAmico:2017mtc}.\footnote{For global fits to all available observables, including fits with combinations of operators, see e.g.\ \cite{Capdevila:2017bsm,Altmannshofer:2017yso}.} The uncertainty in the Wilson coefficient is about $20\%$ to $30\%$. In our model with $\mathcal{L}_{Z'}$, the dominant contribution to this operator is a tree level exchange of $Z'$, yielding
\begin{align}
G_{b_L\mu_L}&=-\frac{C_{Z'bs}C_{Z'\mu\mu}}{M_{Z'}^2} .\label{eq:operator-Gbmu}
\end{align}
It has been shown in \cite{Capdevila:2017bsm,Altmannshofer:2017yso} that adding an operator with electrons instead of muons does not significantly change the requirement for $G_{b_L\mu_L}$. We will therefore ignore such an operator in the following. Note that one can keep the $b\to se^{+}e^{-}$ at the same level as in the Standard Model by suppressing the $C_{Z'ee}$ coupling, e.g.\ by taking $(s_{14}^{eL})^{2}=3/5$.

 We consider now whether the $\mathrm{SO}(10)$ model in question can give a large enough contribution to the $b\to s\mu\mu$ process in order to explain the $\RKK$ anomalies. To answer this, we need to study bounds on other processes, where the $C_{Z'bs}$ and $C_{Z'\mu\mu}$ couplings will be important. It turns out that the mass difference in the $B^0_s$-$\bar{B}^{0}_s$ mixing will already provide a severe constraint.

All processes of the type $b\bar{s}\to s\bar{b}$ contribute to diagrams mixing the $B^0_s$ and $\bar{B}^{0}_s$ mesons. The difference of the masses of the two mass eigenstates is computed in the Standard Model, but with a relatively large theoretical uncertainty (mostly from computation of lattice parameters): $\Delta m_s=(17.4\pm 2.6)\mathrm{ps}^{-1}$~\cite{Bhattacharya:2016mcc}. The experimentally measured value is given much more precisely to be $\Delta m_s=(17.757 \pm 0.021)\mathrm{ps}^{-1}$~\cite{Amhis:2014hma}. Assuming the accurate SM contribution in the low range of the given interval, the largest possible contribution from BSM physics can then be estimated.

In a $Z'$ model, $B$ mixing will be induced by a tree level exchange of the $Z'$ involving two $Z'bs$ couplings.  Following Table I in~\cite{Cline:2017lvv}, and translating our notation to their notation for the vertex $Z'bs$ by $C_{Z'bs}\to g_q s_\theta c_\theta$, the constraint from $B^{0}_s$-$\bar{B}^{0}_s$ mixing reads
\begin{align}
C_{Z'bs}^2\;\mathrm{TeV}^2/M_{Z'}^2 \lesssim 2\cdot 10^{-5}.\label{eq:BB-mixing-bound}
\end{align}
Expressing  $C_{Z'bs}$ from Eq.~\eqref{eq:operator-Gbmu} and inserting into Eq.~\eqref{eq:BB-mixing-bound} gives
    \begin{align}
    (C_{Z'\mu\mu})^{-2}\;M_{Z'}^{2}\;\mathrm{TeV}^{2}\;G_{b_L\mu_L}^{2}=\tfrac{40}{9g'^{2}}M_{Z'}^2\;\mathrm{TeV}^2\;G_{b_L\mu_L}^{2}&\lesssim 2\cdot 10^{-5}.\label{eq:BBmixing-bounds}
    \end{align}

For the $g'$, we can make a simple estimate by noting that the $1$-loop RGE equations for a generic gauge coupling $g$ are $\tfrac{dg}{dt}=\tfrac{\beta_g g^3}{(4\pi)^2}$. We compute the $\beta$ coefficients for the $g$ and $g'$ couplings of the $\mathrm{U}(1)_Y$  and $\mathrm{U}(1)_X$ groups, respectively, to be $\beta_{g}=53/5$ and $\beta_{g'}=77/5$. This holds in the GUT normalization, in which the couplings unify at $M_{\text{GUT}}$, and the particle content is that of the MSSM with additional vector-like states, as listed in Table~\ref{tab:funfields}. Since the $\beta_{g'}>\beta_g$, and $g=g'$ at the GUT scale (we assume the threshold corrections at the GUT scale for $g$ and $g'$ do not differ much), $g'<g$ at low scales, which increases the left-hand side of Eq.~\eqref{eq:BBmixing-bounds} compared to just replacing the value $g'$ with $g$.

If we assume $g' \approx g=0.46$ (best case scenario for satisfying the bound), $M_{Z'}\approx 4\,\mathrm{TeV}$ and $G_{b_L\mu_L}=G_{b_L\mu_L}^{\mathrm{exp}}=-(31\,\mathrm{TeV})^{-2}$, we get $3.6\cdot 10^{-4}$ on the left-hand side of Eq.~\eqref{eq:BBmixing-bounds}, which is too big by one order of magnitude. We could recover to near the allowed bound by taking $M_{Z'}=1\,\mathrm{TeV}$, but that is excluded by the lower bound from LHC data of roughly $4\,\mathrm{TeV}$~\cite{Osland:2017ema,Aaboud:2017buh}.

Therefore to satisfy the bound, at least one order of magnitude suppression has to come from $G_{b_L\mu_L}$. We can obtain a rough estimate for $g'$ by using the previously stated RGE equations for $g$ and $g'$ between the scales $M_{SUSY}\approx M_{Z'}$ and $M_{\text{GUT}}\approx 2\cdot 10^{16}\,\mathrm{GeV}$, assuming unification $g\approx g'$ at $M_{\text{GUT}}$ and the low energy value $g=0.46$. At the scale of $M_{Z'}$, we get $g'^2\approx 0.7 g^2$. This allows us to write the estimate for the operator coefficient $G_{b_L\mu_L}$ as
\begin{align}
G_{b_L\mu_L}\lesssim -\frac{1}{(70.0\,\mathrm{TeV})^2}\,\left(\frac{4\,\mathrm{TeV}}{M_{Z'}}\right)\,\left(\frac{g'}{0.46\;\sqrt{0.7}}\right),
\end{align}
\noindent
which is not sufficient to fully explain the $\RKK$ anomaly measurement, since the mass scale of $70\,\mathrm{TeV}$ is equivalent to the Wilson coefficient value of about $C^{\textrm{BSM}}_{b_L\mu_L}=-0.26$.

We now turn to the general case, in which mixing angles with the $4$th vector-like state other than $\theta^{eL}_{14}$ and $\theta^{dL}_{34}$ are switched on. The particular forms of couplings in Eq.~\eqref{eq:CZbs-coupling}--\eqref{eq:CZee-coupling} now change, but the derivation of the general form in Eq.~\eqref{eq:BBmixing-bounds}, before the coupling $C_{Z'\mu\mu}$ is inserted, still holds. We see that the left-hand side of that inequality is proportional to $(C_{Z'\mu\mu})^{-2}$, so the bound can be reached only if the absolute value of the coupling gets enhanced. Direct computation via Eq.~\eqref{eq:mixing-angle-parametrization}, in the small $\theta^{eL}_{12}$,$\theta^{eL}_{13}$,$\theta^{eL}_{23}$ angle limit, shows the value of the coupling $C_{Z'\mu\mu}$ in the general case to be
\begin{align}
C_{Z'\mu\mu}&=\tfrac{g'}{2\sqrt{10}}\Big((-3)\;(c^{eL}_{24})^{2}+\left((-3)(s^{eL}_{14})^{2}+2(c^{eL}_{14})^{2}\right)(s^{eL}_{24})^{2}\Big).
\end{align}
The coupling does not depend on $\theta^{eL}_{34}$. Furthermore, for any values of angles $\theta^{eL}_{14}$ and $\theta^{eL}_{24}$, the second factor is between $-3$ and $2$. The former value is reached for the already considered special case $\theta^{eL}_{24}=0$, while the latter is achieved for $\theta^{eL}_{14}=0$ and $\theta^{eL}_{24}=\pi/2$. This shows that the absolute value of the coupling $C_{Z'\mu\mu}$ reached its maximum already in the special case we considered earlier and cannot be enhanced, implying that the bound in Eq.~\eqref{eq:BBmixing-bounds} is violated in the general case at least as much as it was before.

The above considerations show that fulfilling the bound from the mass split in $B_s^0$-$\bar{B}^0_s$ mixing and explaining the full size of the $\RKK$ anomalies is not possible in our model without tension in the mass of the $Z'$ coming from LHC bounds. If we abandon the explanation of the $\RKK$,
then the angle $\theta^{dL}_{34}$ can be taken small enough to suppress the $ C_{Z'bs}$ coupling as much as needed, and thus avoid the stringent experimental bounds from $B_s^0$-$\bar{B}^0_s$ mixing. This is the path we shall follow in the remainder of the paper.

\subsection{Non-universal $Z'$ and charged lepton flavour violating processes\label{section:pheno-mue}}

There are various BSM processes that are induced in our model due to the presence of both the $Z'$ and vector-like states. A full phenomenological analysis is beyond the scope of this paper, not least because we consider an entire class of models and the Yukawa sector in the quark and charged lepton sectors needs to be appended, as discussed at the start of section~\ref{section:constraints}. However, the question of charged lepton flavour violating processes in $\mathrm{SO}(10)$ GUT models under consideration is a very fascinating one, since they offer 3 predictive advantages over ``bottom-up'' extensions of the SM with $Z'$ and vector-like fermions added as unrelated extra ingredients:

\begin{enumerate}
\item In  ``bottom-up'' extensions of the SM, the gauge coupling $g'$ related to the gauge boson $Z'$ can in principle take any value. It can, for example, be arbitrarily small, so that the flavour-violating effects are negligibly small. In an $\mathrm{SO}(10)$ model, the $Z'$ unifies with the other interactions at the GUT scale; unification conditions thus give the value of the coupling to be $g'\approx 0.46\,\sqrt{0.7}$, as already discussed in Section~\ref{section:pheno-RKK}. 
\item The $Z'$-charges of fermions (both of the first 3 families, as well as of the extra vector-like family) are arbitrary in a SM based model. In an $\mathrm{SO}(10)$ model, these charges are automatically determined by the choice of $\mathrm{SO}(10)$ representations for the fermions. 
\item The couplings of the SM fermions to the $Z'$ depend on the mixing between the  first 3 families and the vector-like one. In an $\mathrm{SO}(10)$ model, the Yukawa matrices in the quark and lepton sectors are related, so the mixing is related.
\end{enumerate}

The above considerations show that in any specific model, when the Yukawa matrices are known, any flavour violating process can be computed, since the $g'$ and the associated fermion charges are known. Furthermore, it shows that the underlying $\mathrm{SO}(10)$ symmetry can relate flavour violating processes from different sectors. The original motivation of the $\mathrm{SO}(10)$ models with a low-energy $Z'$ and vector-like states was an explanation of the $\RKK$ anomalies; this explanation was shown to be disfavored in Section~\ref{section:pheno-RKK}, but the considerations in flavour violating processes by themselves serve as an excellent motivation for such models in this paper and beyond.

As a demonstration, we investigate in this section two flavour violating processes for this class of $\mathrm{SO}(10)$ models: the lepton flavour violating decay $\mu \to 3e$ and the $\mu$-$e$ conversion in a muonic atom, and show how they are related. In particular, we investigate the branching ratios $\mathrm{Br}(\mu \to 3e)$ and $\mathrm{Br}(\mu^-\, N \rightarrow e^-\, N)$, where $N$ is the nucleus of the muonic atom.

The choice for these particular BSM processes is manifold: first, they are mediated by a tree-level exchange of the $Z'$, ensuring a relatively large contribution from BSM physics. Second, they involve first two generation particles, with sensitive experiments designed to look for these processes. Third, these two processes are sensitive to the mixing between the vector-like and the chiral states in the charged lepton sector and to the charged lepton mixing angle $\theta_{12}^{eL}$, which is typically the dominant charged lepton mixing contribution to the PMNS matrix, especially in an $\mathrm{SO}(10)$ model where the charged lepton mixing angles are connected to the quark ones.\footnote{This can further have an impact on the PMNS mixing angles and the Dirac CP phase $\delta_{\text{PMNS}}$ via lepton mixing sumrules, cf.\ \cite{sumrule}.}

For the purposes of this demonstration, we make the assumptions that are outlined in the beginning of Section~\ref{section:constraints}; although we do not assume the exact form of Yukawa matrices from Eq.~\eqref{eq:ude_mass_matrix}, we do assume that the possible modifications correcting the bad $\mathrm{SU}(5)$ relations in a complete model do not change the structure of the matrices appreciably (the entries do not change their order of magnitude for example). More specifically, we take that the relations in Eq.~\eqref{eq:theta5}--\eqref{eq:theta4-2} hold at least approximately:
there is negligible mixing in $\theta^{qL}_{i5}\approx\theta^{qR}_{i5}\approx 0$ (for $i=1,2,3$ and $q=u,d,e$), while the $i$-$4$ angles can take in principle any value, but their sizes are related according to Eq.~\eqref{eq:theta4-1} and \eqref{eq:theta4-2}.

\begin{figure}[htb]
  \minipage{0.5\textwidth}
  \centering
  \includegraphics[width=0.8\textwidth]{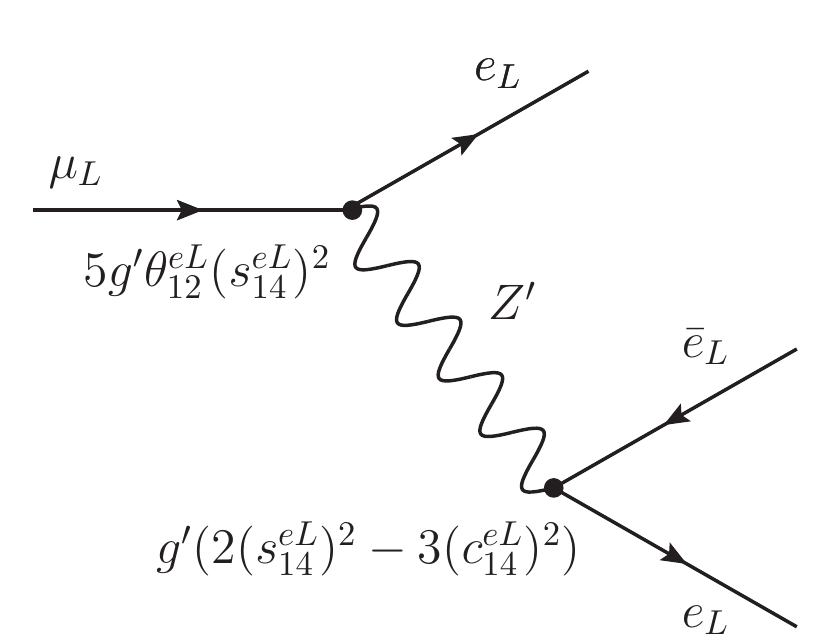}
  \endminipage
  \minipage{0.5\textwidth}
  \centering
  \includegraphics[width=0.8\textwidth]{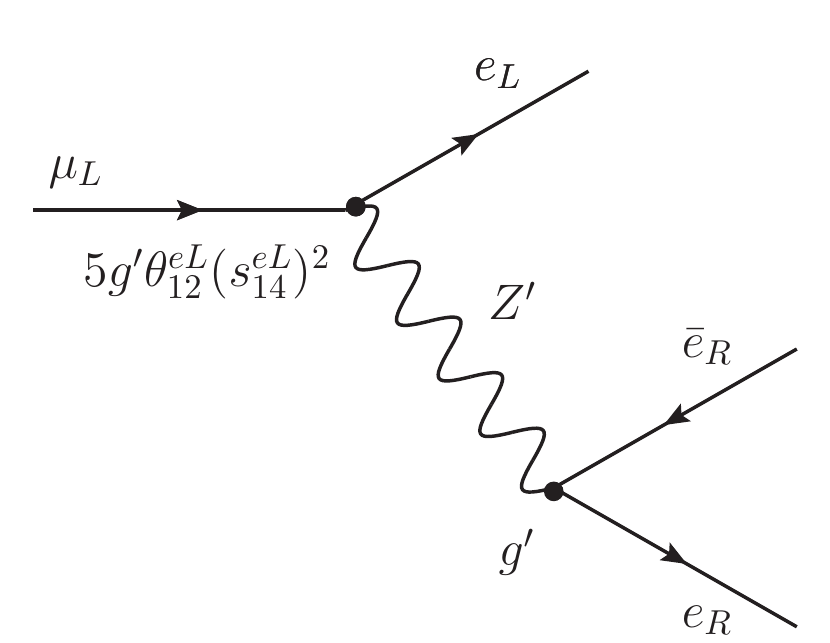}
  \endminipage
  \caption{Lepton flavour violating decay $\mu \to 3e$ via the $Z'$. On the left-hand side the $Z'$ couples to $\bar{e}_L e_L$, whereas on the right-hand side it couples to $\bar{e}_R e_R$. In both diagrams the couplings of the left- and right-handed leptons to the $Z'$ have to be normalized with a factor of $(2\sqrt{10})^{-1}$. Furthermore, $\theta_{12}^{eL} \ll 1$ is assumed.}
  \label{fig_mu3e_diagram}
\end{figure}

\begin{figure}[htb]
  \minipage{0.5\textwidth}
  \centering
  \includegraphics[width=0.9\textwidth]{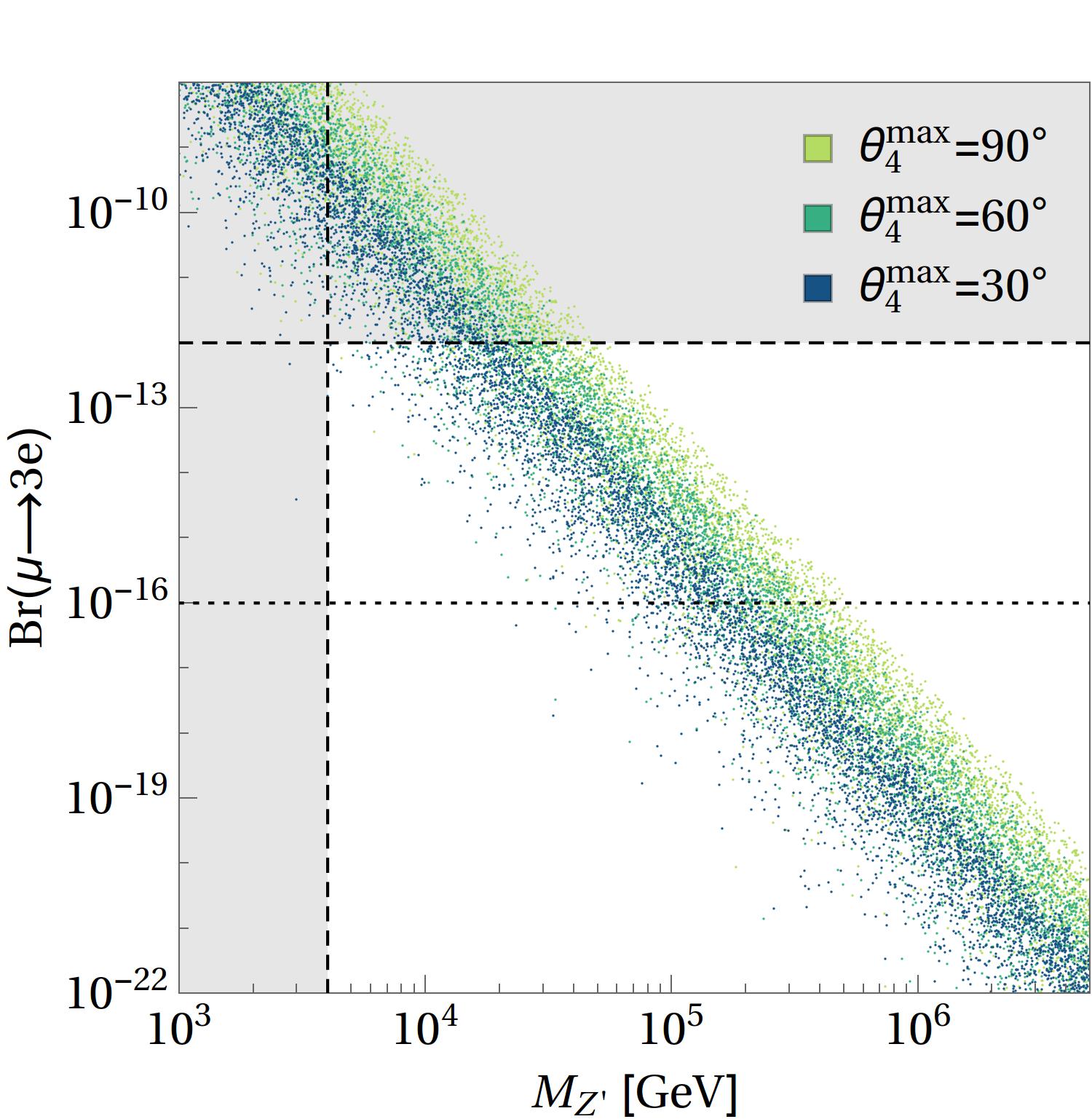}
  \endminipage
  \minipage{0.5\textwidth}
  \centering
  \includegraphics[width=0.9\textwidth]{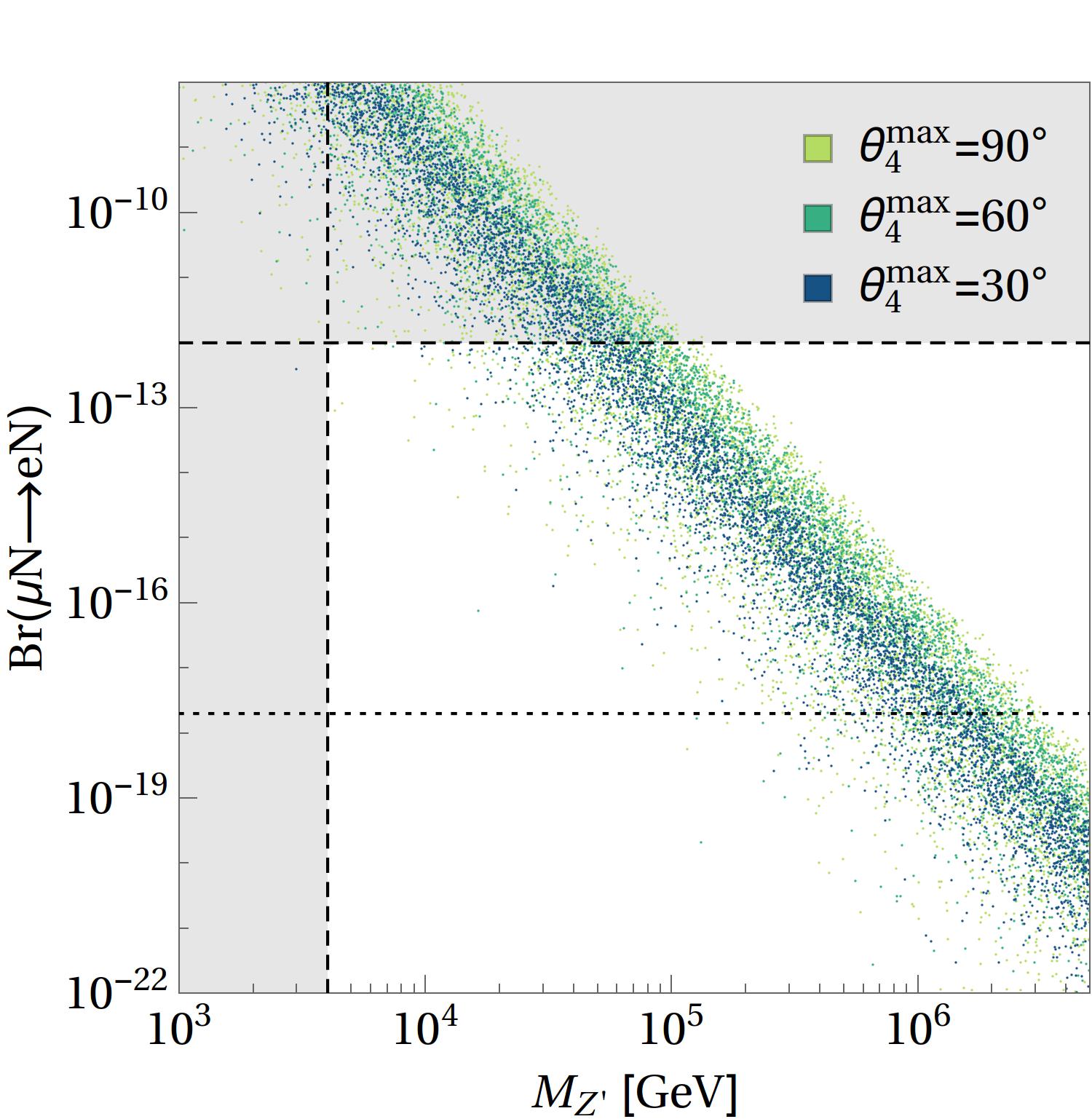}
  \endminipage
  \caption{The branching ratio of $\mu\to 3e$ (left) and $\mu$-$e$ conversion (right) depends on the mass scale of the $Z'$, as shown in these log-log scatter plots, where the other parameters are varied (see main text). The grey areas represent regions excluded by LHC bounds on $Z'$ mass ($M_{Z'}>4\mathrm{TeV}$), and the branching ratio  bounds come from experiments dedicated to measuring flavour violating processes. The other dotted horizontal lines represent the estimated reach of future $\mu\to 3e$ and $\mu$-$e$ conversion experiments.
   \label{fig_mue_plot}
   }
\end{figure}

In a scenario when $\theta_{12}^{eL}\gg \theta_{12}^{eR}$ and $\theta_{14}^{eL}\gg \theta_{24}^{eL},\theta_{34}^{eL}$, the lepton flavour violating decay \hbox{$\mu \to 3e$} is realized dominantly via the diagrams shown in Figure~\ref{fig_mu3e_diagram}. If $\theta_{24}^{eL}$ is non-negligible, the interaction couplings become more complicated expressions, and when $\theta_{12}^{eR}$ is engaged, analogous processes with right-handed fermions in the upper fermion line of the diagram contribute. The full contribution to the branching ratio in the most general case can be computed with eq.~(21) in \cite{Langacker:2008yv}, where the required parameters are the general form of the $Z'$ couplings, the $Z'$ mass and the $Z'$-$Z$ mixing parameter $\theta$, which is computed from the gauge boson mass matrix. We take the realistic case $\theta\sim M_{Z}^2/M_{Z'}^2$. The $Z'$ couplings are computed in the flavour basis of fermions from the known $\mathrm{U}(1)_X$ charges of the fermions, in which $Z'$ is flavour diagonal, and then rotated to the mass eigenbasis of fermions using the convention of mixing angles from Eq.~\eqref{eq:mixing-angle-parametrization}; the analytic expression is prohibitively long and we omit it here. The general formula is used for the computation of $\mbox{Br}(\mu \to 3e)$ in the left scatter plot of Figure~\ref{fig_mue_plot}, which we shall describe in more detail later. As a rough approximation of the general formula, in the case when $\theta\approx 0$, $\theta_{12}^{eR}\ll\theta_{12}^{eL} \ll 1$ and $\theta_{14}^{eL}\gg\theta_{24}^{eL}$, we obtain 

\begin{align}
  \mathrm{Br}(\mu \rightarrow 3e) &\approx 2.27 \cdot 10^{-11} \left(\frac{\theta_{12}^{eL}}{3^\circ}\right)^2 \left(\frac{4\,\mathrm{TeV}}{M_{Z'}}\right)^4 \left(\frac{g'}{0.46\,\sqrt{0.7}}\right)^4 (s_{14}^{eL})^4 \Big[ 1 + 2 \big( 3(c_{14}^{eL})^2 - 2(s_{14}^{eL})^2 \big)^2 \Big].\label{eq:Br-mu3e-simplified}
\end{align}

\noindent
This can be compared with the experimental bound on $\mbox{Br}(\mu \to 3e)$, which is $1.0 \cdot 10^{-12}$~\cite{Agashe:2014kda}. 

\begin{figure}[htb]
  \centering
  \includegraphics[width=0.4\textwidth]{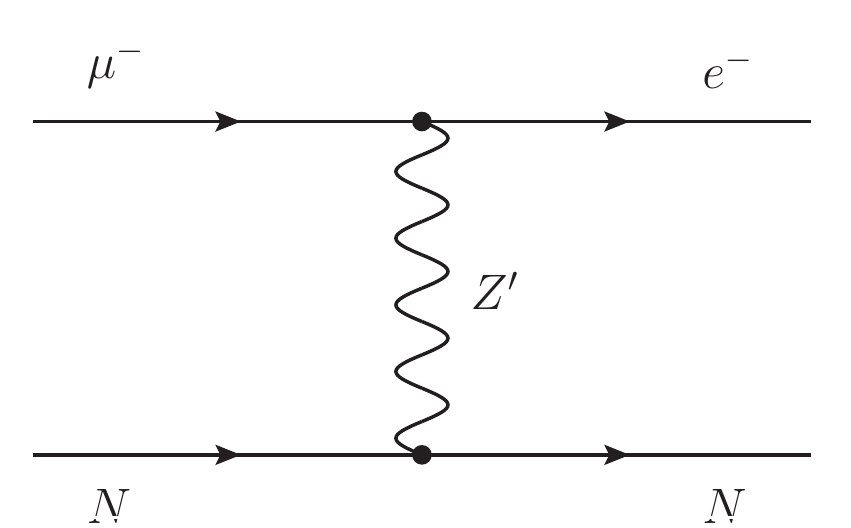}
  \caption{Schematic diagram for the $\mu$-$e$ conversion in a muonic atom. In the diagram, in order to get the different processes contribution to the $\mu$-$e$ conversion, $N$ has to be replaced by left- and right-handed up and down quarks, with the specific couplings to the $Z'$.}
  \label{fig_mue_diagram}
\end{figure}

The dominant process for $\mu$-$e$ conversion in a muonic atom in the scenario $\theta_{12}^{eR}\ll\theta_{12}^{eL}$ and $\theta_{14}^{eL}\gg\theta_{24}^{eL}$ proceeds via the diagram shown in Figure~\ref{fig_mue_diagram}. In our analysis, we will consider a gold ($\mathrm{Au}$) atom. We compute $\mathrm{Br}(\mu^-\, \mathrm{Au} \rightarrow e^-\, \mathrm{Au})$ for the general case of mixing angles from equation~(22) in \cite{Langacker:2008yv}, with $Z$-$Z'$ mixing $\theta\sim M_{Z}^2/M_{Z'}^2$. This computation is used in the right scatter plot of Figure~\ref{fig_mue_plot}, with the parameters related to the nucleus of $\mathrm{Au}$ atoms taken from~\cite{Kitano:2002mt}. In the case when $\theta\approx 0$, $\theta_{12}^{eR}\ll\theta_{12}^{eL} \ll 1$ and $\theta_{14}^{eL}\gg\theta_{24}^{eL}$, the general formula can again be simplified:

\begin{align}
  \mathrm{Br}(\mu^-\,\mathrm{Au} \rightarrow e^-\,\mathrm{Au}) &\approx 1.51 \cdot 10^{-8} \left(\frac{\theta_{12}^{eL}}{3^\circ}\right)^2 \left(\frac{4\,\mathrm{TeV}}{M_{Z'}}\right)^4 \left(\frac{g'}{0.46\,\sqrt{0.7}}\right)^4 (s_{14}^{eL})^4.\label{eq:Br-mue-simplified}
\end{align}

\noindent
The experimental bound on $\mathrm{Br}(\mu^-\, \mathrm{Au} \rightarrow e^-\, \mathrm{Au})$ is given by $7.0 \cdot 10^{-13}$~\cite{Agashe:2014kda}.

We emphasize that the simplified expressions for the branching rations of \hbox{$\mu\to 3e$} and \hbox{$\mu\,\mathrm{Au}\to e\,\mathrm{Au}$} in Eq.~\eqref{eq:Br-mu3e-simplified} and \eqref{eq:Br-mue-simplified} hold in the case $\theta\approx 0$, $\theta_{12}^{eR}\ll\theta_{12}^{eL} \ll 1$ and $\theta_{14}^{eL}\gg\theta_{24}^{eL}$; this includes any $Z'$ model with vector-like fermions whose $Z'$ charges are the $\mathrm{U}(1)_X$ charges from Table~\ref{tab:funfields}, regardless on whether the theory has an underlying $\mathrm{SO}(10)$ symmetry; one of the advantage of the $\mathrm{SO}(10)$ class of models however is that we know the value of $g'$. The numerical prefactors in Eq.~\eqref{eq:Br-mu3e-simplified} and \eqref{eq:Br-mue-simplified} are above the experimental bounds for these processes; given a CKM-like angle for $\theta_{12}^{eL}$, $s_{14}^{eL}$ has to be much smaller than $1$ and/or the $Z'$ mass has to be much larger than $4\,\mathrm{TeV}$.

We perform a parameter scan over the mixing angles and compute the branching ratios using the general formulas for the two processes from \cite{Langacker:2008yv}. We assume the following reasonable scenario:
\begin{itemize}
\item The branching ratio can be arbitrarily reduced by reducing the mixing angles, such as the $\theta_{12}^{eL}$ angle. We assume, however, CKM-like mixing angles in the lepton sector. Furthermore, we assume most of the CKM angles are coming from the down sector, and we know empirically that the $\theta_{13}^\text{CKM},\theta_{23}^\text{CKM}\ll \theta_{12}^\text{CKM}$. We assume the same features to be present in the right mixing angles, since the $y_1$ $3\times 3$ matrix in Eq.~\eqref{eq:ude_mass_matrix} is symmetric. All of the above considerations are combined into taking the following values for the mixing angles between the chiral families:
	\begin{align}
	\theta_{13}^{qX},\theta_{23}^{qX}&:\quad 0,\\
	\theta_{12}^{eX},\theta_{12}^{dX}&:\quad r\cdot \theta_{12}^\text{CKM},\\
	\theta_{12}^{uX}&:\quad 0.2\,r\cdot \theta_{12}^\text{CKM},
	\end{align} 
where $\theta_{12}^\text{CKM}=0.227$, the labels $X$ and $q$ run over $X=L,R$ and $q=u,d,e$, and finally $r\in[0.8,1.0]$ is a uniformly chosen random number different for each of the parameters, since we do not assume Eq.~\eqref{eq:ude_mass_matrix} to hold exactly.
\item For the chiral-vector mixing angles, we assume that Eq.~\eqref{eq:theta5}--\eqref{eq:theta4-2} approximately hold. Also, the $\theta_{34}$ angles do not come into the branching ratio formulas, since there is no $1$-$3$ and $2$-$3$ mixing. This considerations amount to the following values for the chiral-vector angles:
	\begin{align}
	\theta_{i5}^{qX}&:\quad 0,\\
	\theta_{j4}^{uL}, \theta_{j4}^{uR}, \theta_{j4}^{dL}, \theta_{j4}^{eR}&:\quad p\cdot\kappa_{j},\\
	\theta_{j4}^{dR}, \theta_{j4}^{eL}&:\quad p\cdot\lambda_j,\\
	\theta_{34}^{qX}&:\quad \text{irrelevant},
	\end{align}
	where $i=1,2,3$, $j=1,2$, the labels $X$ and $q$ run over $X=L,R$ and $q=u,d,e$. Also, $p$, $\kappa_j$ and $\lambda_j$ are randomly chosen (including for each $j$) uniformly distributed values in the following ranges: $p\in[0.5,1.0]$, $\kappa_j,\lambda_j\in [0,\theta_{4}^{max}]$. In the scattering plots we use three different ensembles of points based on different values of the maximum $j$-$4$ angles $\theta_{4}^{\text{max}}$. The scales $\kappa_j$ and $\lambda_j$ for the $j$-$4$ mixing angles are intended to represent
		\begin{align}
		\kappa_j&\sim \arctan\left((Y_3)_j\bar{v}_R/m_{45}\right),\\
		\lambda_j&\sim \arctan\left((Y_2)_j v_R/m_{10}\right), 
		\end{align}  
		where $(Y_3)_j$ and $(Y_{2})_j$ are the $j$-th component of the vectors of parameters $Y_3$ and $Y_2$, respectively.
\item We fix the gauge couplings to be $g=0.46$ and $g'=0.46\sqrt{0.7}$ based on unification constraints, see Section~\ref{section:pheno-RKK}. Also, we take the $Z$-$Z'$ mixing parameter to be exactly $\theta=M_Z^2/M_{Z'}^2\ll 1$. 
\end{itemize}

The results are the scatter plots of the $\mathrm{Br}(\mu \to 3e)$ and $\mathrm{Br}(\mu^-\, \mathrm{Au} \rightarrow e^-\, \mathrm{Au})$ for a given $Z'$ mass, see Figure~\ref{fig_mue_plot}. It should be mentioned here that the branching ratios can in principle drop arbitrarily low for any mass $M_{Z}'$ if the $i$-$4$ mixings with the extra vector-like family are all artificially small, e.g.~if $m_{10},m_{45}\gg v_{R},\bar{v}_{R}$. In the generic part of the parameter space, however, we see a linear-drop with $M_{Z'}$ in the log-log scatter plots. While the lower bound in the scatter plots is not strict, the upper bound with the yellow points is as high as the branching ratio can go at a given $Z'$ mass in our class of models. The shaded areas of these plots are excluded regions based on current experimental bounds for the branching ratios (\hbox{$\mathrm{Br}(\mu \to 3e)<1.0 \cdot 10^{-12}$ and \hbox{$\mathrm{Br}(\mu^-\, \mathrm{Au} \rightarrow e^-\, \mathrm{Au})< 7.0 \cdot 10^{-13}$}~\cite{Agashe:2014kda}}) and the $M_{Z'}$ coming from the LHC ($M_{Z'}>4\,\mathrm{TeV}$~\cite{Osland:2017ema,Aaboud:2017buh}). The other horizontal lines are the sensitivities of future experiments; they can probe \hbox{$\mathrm{Br}(\mu \to 3e)$} with a sensitivity of $10^{-16}$~\cite{Blondel:2013ia}, and the branching ratio of $\mu$-$e$ conversion in a muonic $\mathrm{Ti}$ atom with a sensitivity of $2\cdot 10^{-18}$~\cite{Knoepfel:2013ouy}. We can see from the two plots that the $\mu$-$e$ conversion is the more sensitive of the two experiments for probing our class of $\mathrm{SO}(10)$ models. The future sensitivity of the $\mu$-$e$ conversion experiments has the potential to probe for $M_{Z'}$ masses up to roughly $3000\,\mathrm{TeV}$, much higher than the reach of the LHC or any planned next-generation collider. 

\begin{figure}[htb]
  \centering
  \includegraphics[width=8cm]{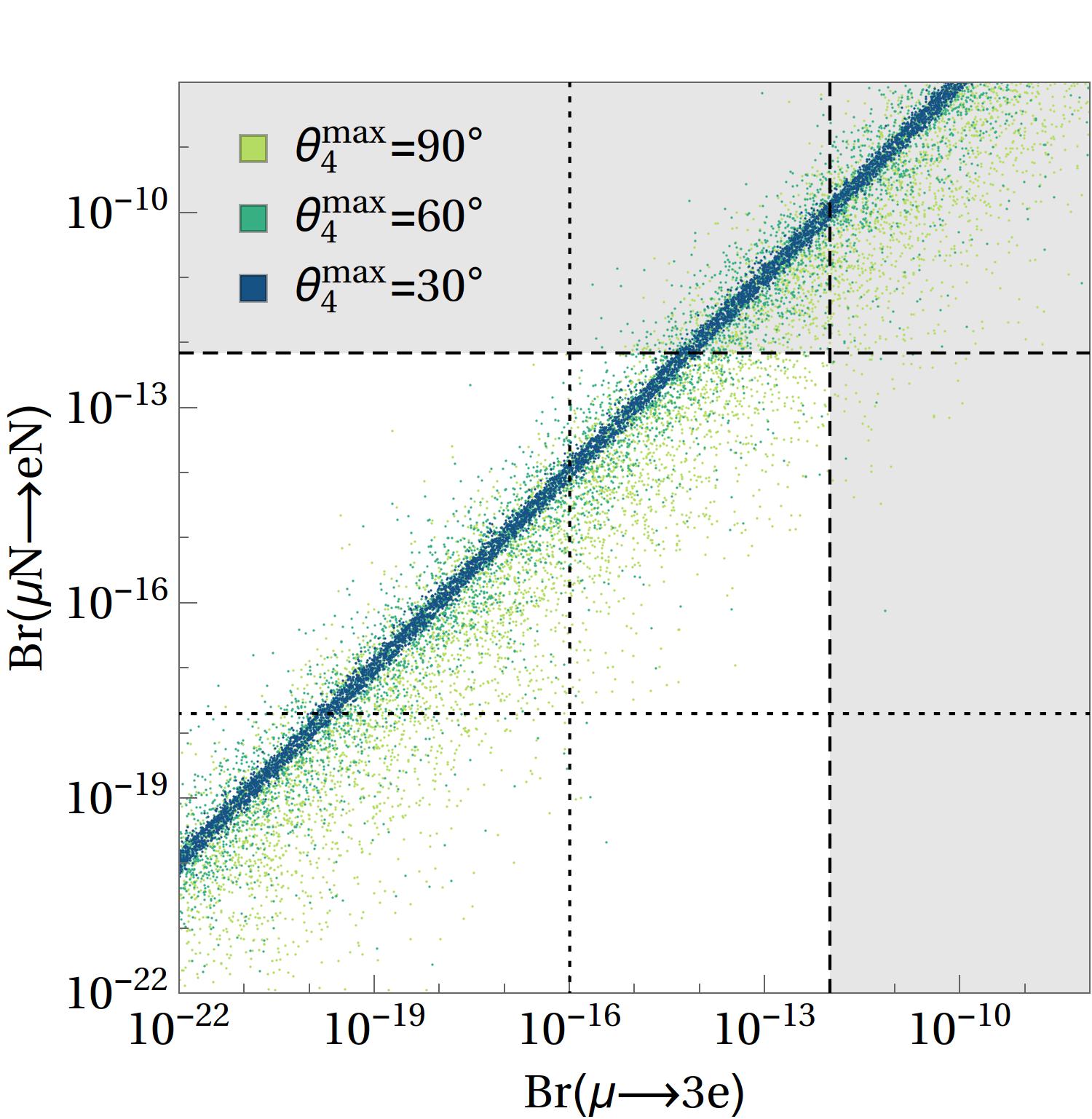}
  \caption{The log-log scatter plot of the branching ratios of $\mu\to 3e$ and $\mu$-$e$ conversion flavour violating processes. We see a strong correlation between the two processes, with the correlation already predicting the value of one branching ratio by knowing another to within factor $3$ if all the chiral-vector $1$-$4$ and $2$-$4$  angles are smaller than $30^\circ$. The grey areas represent experimentally excluded regions, and the other two lines the future experimental sensitivities.
   \label{fig_correlation}
   }
\end{figure}

Another striking feature appears when we compare the two branching ratios for the same parameter values (mixing angles and $M_{Z'}$). A log-log scatter plot is shown in Figure~\ref{fig_correlation}, where the grey areas are again the experimentally excluded regions and the extra lines show future sensitivities. We can see that the two branching ratios are correlated; if the chiral-vector mixing angles are not big ($\theta_{4}^{\text{max}} <30^\circ$ is already sufficient, see blue points), the correlation is constrained to within much less than one order of magnitude. This correlation is actually a feature of the underlying $\mathrm{SO}(10)$ symmetry of our class of models. In a non-$\mathrm{SO}(10)$ model, the processes $\mu\to 3e$ and $\mu$-$e$ conversion are in principle unrelated, due to the relevant couplings of the $Z'$ to charged leptons and quarks being unrelated (the bottom $Z'$ couplings in Figures~\ref{fig_mu3e_diagram} and \ref{fig_mue_diagram}). The scattering plots are valid for the reasonable parameter scenario outlined earlier in this section; this include a more stringent scenario where the Yukawa matrices are exactly those of Eq.~\eqref{eq:ude_mass_matrix}.

The correlation of the $\mu\to 3e$ and $\mu$-$e$ conversion branching ratios is thus a general prediction of the studies $\mathrm{SO}(10)$ class of models. If an experiment measures the $\mu$-$e$ conversion at a certain rate, Figure~\ref{fig_correlation} tells us in which range the measurement of the other flavour violating process $\mu\to 3e$ is expected to occur. Furthermore, a successful measurement of $\mu$-$e$ conversion would put an upper bound on the $M_{Z'}$ (ensemble of yellow points in Figure~\ref{fig_mue_plot}). 

\section{Conclusions}
A $Z'$ gauge boson with mass around the (few) TeV scale can be a fascinating remnant of a Grand Unified Theory. When the $\mathrm{U}(1)'$ gauge symmetry of the $Z'$ is embedded in an underlying GUT group, the latter relates the $Z'$-couplings of the different fermions of the Standard Model (SM). Further testable remnants are TeV scale vector-like states which can arise from additional GUT matter representations. These vector-like states can have an interesting impact on the flavour structure via enlarged fermion mass matrices, and induce non-universal $Z'$ couplings with a rich phenomenology.

Recently, $Z'$ models with non-universal couplings to the SM fermions due to extra vector-like states have received attention as candidates for explaining the present $R_K$, $\RK$ anomalies. As one part of our paper, we have revisited this possibility by further developing consistent $\mathrm{SO}(10)$ GUT models with a non-universal $Z'$ at low energy arising from a spontaneously broken $\mathrm{U}(1)_X$ gauge subgroup and a vector-like fourth family.
The $Z'$ has non-universal couplings due to the mixing between the chiral quark and lepton families and
the vector-like family which couples with non-universal $\mathrm{U}(1)_X$ charges.

However, despite all the ingredients for explaining the $R_K$, $\RK$ anomalies being present in such models, we have shown that they fail to provide a consistent explanation of the anomalies due to the constraints from the $B_s$ mass difference and from LHC bounds on the $M_{Z'}$, when taken into account that the 
maximal value of the coupling of $Z'$ to muons is restricted by $\mathrm{SO}(10)$.

On the other hand, models of this type have a rich phenomenology in the lepton sector: 
Due to the constraints from $\mathrm{SO}(10)$ on the Dirac neutrino Yukawa couplings, we have argued that a conventional seesaw mechanism is not possible with right-handed neutrinos at the $\mathrm{TeV}$ scale (assuming that $\mathrm{U}(1)_X$ is broken at the scale of a few $\mathrm{TeV}$). 

To complete the class of models with a consistent neutrino sector, we have demonstrated that it is possible to implement both a linear seesaw~\cite{Malinsky:2005bi} and an inverse seesaw mechanism~\cite{Dev:2009aw} in $\mathrm{SO}(10)$ with vector-like states and TeV scale $Z'$. Either mechanism (or even a simultaneous combination of the two) can generate the masses of the light neutrinos. Furthermore, it is possible to forbid the inverse seesaw coupling, while the linear seesaw contribution naturally gives the correct light neutrino mass scale. This represents a new example of the linear seesaw mechanism. 

Our class of models can be viewed as a new route towards constructing highly predictive and testable flavour models with a number of compelling phenomenological aspects. For example, with the right-handed (sterile) neutrinos in reach of collider experiments, this opens up another interesting window towards testing the seesaw mechanism in GUTs (see e.g.\ \cite{Antusch:2016ejd,Deppisch:2015qwa} and references therein). Moreover, we considered the flavour violating processes of $\mu \to 3e$ decay and $\mu$-$e$ conversion 
in a muonic atom; we showed that within a reasonably general parameter scenario, there is a strong correlation between the two processes due to the underlying $\mathrm{SO}(10)$ symmetry. We also considered present and future experimental sensitivities for probing this class of models. We showed that future sensitivities of $\mu$-$e$ conversion experiments will have an experimental reach of $Z'$ masses up to about $3000\,\mathrm{TeV}$, much higher than any planned future collider.

In conclusion, we have clarified several issues regarding $\mathrm{SO}(10)$ GUTs with a non-universal $Z'$ at low energy due to a vector-like fourth family: On the one hand, we have shown that models of this type cannot provide a consistent explanation of the present $R_K$, $\RK$ anomalies. On the other hand, we have shown how a consistent neutrino sector can be constructed for such models, requiring right-handed (sterile) neutrino masses at the mass scale of the $Z'$ (or below). Furthermore, we showed the predictive potential of such models by relating flavour violating processes involving interactions from different fermionic sectors. The resulting models thus feature a rich phenomenology and may be tested with high precision at present and future colliders and in lepton flavour violating experiments.

\section*{Acknowledgements}
This work has been supported by the Swiss National Science Foundation.
S.\,F.\,K. acknowledges the STFC Consolidated Grant ST/L000296/1 and the European Union's Horizon 2020 Research and Innovation programme under Marie Sk\l{}odowska-Curie grant agreements Elusives ITN No.\ 674896 and InvisiblesPlus RISE No.\ 690575.

\appendix

\section{Doublet-triplet splitting\label{A}}
Since the implementation of the linear seesaw mechanism depends on the value of the doublet VEV $\bar{v}_L$ in $\mathbf{\overline{16}}_H$, the details of the doublet-triplet splitting are crucial to the implementation of that mechanism.

Before $\mathrm{U}(1)_X$-breaking, the low energy MSSM Higgs doublets arise purely
from the doublets in the $\mathbf{10}_H$, and the Higgs doublets from the $\mathbf{16}_H$, $\overline{\mathbf{16}}_H$
have GUT scale masses. However, when the $\mathrm{U}(1)_X$  is broken at the few $\mathrm{TeV}$ scale there will be a mixing induced between
the GUT scale Higgs doublets from the $\mathbf{16}_H$, $\overline{\mathbf{16}}_H$ and the light Higgs doublets
from the $\mathbf{10}_H$ of different $\mathrm{U}(1)_X$  charges. This is a very small but crucial effect, which allows to explain the small neutrino masses of the right order of magnitude via the linear seesaw mechanism. To clarify this point, we write down the mass matrix $\mathbf{M}_D$ for the doublets with $Y/2$ charges $\pm 1/2$:

\begin{align}
\mathbf{M}_D&=
    \begin{pmatrix}
    m^{H}_{10}-3\eta' \langle \mathbf{54}_H\rangle&\lambda v_R\\
    -\bar{\lambda}\bar{v}_R&\eta\langle \mathbf{45}_H\rangle_1\\
    \end{pmatrix}.\label{eq:mass-doublets}
\end{align}

This matrix is written in the basis
\begin{align}
\mathbf{D}&=(D_{10},D_{\overline{16}})^T,&\mathbf{\overline{D}}&=(\overline{D}_{10},\overline{D}_{16})^T,\label{eq:definition-doublets}
\end{align}
coming into the superpotential as
\begin{align}
W&=\mathbf{D}^T\,\mathbf{M}_D\,\mathbf{\overline{D}}+\ldots\,.
\end{align}
The vectors of states $D$ and $\overline{D}$ denote the hypercharge $+1/2$ and $-1/2$ doublets, respectively, while their index denotes the representation where they are located.

In order to achieve doublet-triplet splitting in this example, as seen from Eq.~\eqref{eq:mass-doublets}, we introduced the representations $\mathbf{54}_H$ and $\mathbf{45}_H$ in the Higgs sector (even matter parity). The former enables  to tune the mass of the doublet to be low in the $(1,1)$ entry, while the latter, assuming its VEV is aligned in the $\mathrm{SU}(5)$ singlet direction, replaces the direct mass term $\mathbf{16}_H\,\mathbf{\overline{16}}_H$, which is forbidden by the global charges in our models of Table~\ref{table:global-charges}. The mixing between the Higgs doublets is due to the terms $\mathbf{10}_H \cdot \mathbf{16}_H \cdot \langle \mathbf{16}_H \rangle_R$ and $\mathbf{10}_H \cdot \overline{\mathbf{16}}_H \cdot \langle \overline{\mathbf{16}}_H \rangle_R$ with respective couplings $\lambda$ and $\bar{\lambda}$ in the superpotential, where the angle brackets indicate the VEVs at the $\mathrm{TeV}$ scale. Below the $\mathrm{U}(1)_X$ breaking scale but above the electroweak breaking scale the mixing induces a relative admixture of the $\mathbf{16}_H,\overline{\mathbf{16}}_H$ doublets into the MSSM Higgses roughly of the order of $v_R/M_{\text{GUT}}$.

We assume that other doublets from the Higgs sector do not couple with the block in Eq.~\eqref{eq:mass-doublets}, and thus do not interfere in the mechanism.
In Table~\ref{table:global-charges}, Example I gives exactly the doublet mass matrix as that in Eq.~\eqref{eq:mass-doublets}. In all three examples of the table, the $2\times 2$ block form is justified, since the other fields introduced in the Higgs sector either do not have any doublets of the type $(1,2,\pm 1/2)$ (the representations $\mathbf{45}_H$ and $\mathbf{54}_H$), or there is no fermionic counterpart field to couple them to the $\mathbf{10}_H$ or $\mathbf{16}_H$ states in the $2\times 2$ block (holds true for $\mathbf{210}_{H}$).

Given the form in Eq.~\eqref{eq:mass-doublets}, the doublet in the $\overline{\mathbf{16}}_H$ acquires a VEV $\bar{v}_L$ of the order of the EW scale times the small admixture with the $\mathbf{10}_H$, i.e.\
\begin{align}
\label{eq_linear_seesaw_2}
	\bar{v}_L & \sim v_u\,\frac{\lambda v_R}{\eta\,\langle\mathbf{45}_H\rangle_1}.
\end{align}
Plugging this into Eq.~\eqref{eq_linear_seesaw} shows that the $\mathrm{TeV}$ scale dependence drops out\footnote{$D$-flatness is achieved by taking the VEVs of the $\mathbf{16}_H$ and $\mathbf{\overline{16}}_H$ to be equal, i.e.~$v_R=\bar{v}_R$, which means their ratio is equal to $1$ exactly.} and the light neutrino masses are
\begin{align}
\label{eq_linear_seesaw_3}
	m_\nu & \sim y_1v_u\,\frac{\bar{v}_L}{\bar{v}_R} \sim (y_1\lambda^2/\eta^2)\,\frac{v_u^2}{\langle\mathbf{45}_H\rangle_1} \;,
\end{align}
yielding the usual ratio from the high scale seesaw mechanism of $v_u^2/M_{\text{GUT}}$, automatically giving the correct scale for left-handed neutrinos.\footnote{Getting the correct neutrino mass scale actually demands that the seesaw scale is a few orders of magnitude below the GUT scale. In our case, we can achieve the same effect by choosing $\lambda$ and $\eta$ appropriately.} Notice that the relevant admixture for linear seesaw is controlled by the $\lambda$ parameter, and not $\bar{\lambda}$.

We emphasise that the mixing between the doublets in the $\mathbf{10}_H$ and in the $\mathbf{\overline{16}}_H$ or $\mathbf{16}_H$ is always controlled by $\mathrm{U}(1)_X$ breaking, whatever the detailed mechanism of providing entries of correct scale in $M_D$ may be.

\section{Models with exclusive mechanisms of neutrino mass generation\label{Appendix:models}}
\subsection{Example II: linear-seesaw only\label{Appendix:II}}

We describe here the details of the model under Example II in Table~\ref{table:global-charges}. Here, we introduce into the Higgs sector (even matter parity fields) beside the $\mathbf{10}_H$, $\mathbf{16}_H$ and $\mathbf{\overline{16}}_H$ also the $\mathbf{210}_H$, $\mathbf{45}_H$, and two copies with different global charges of the $54$-dimensional representations: $\mathbf{54}_H$ and $\mathbf{54}'_H$.

This model is constructed so that the neutrinos get masses only from linear seesaw, while the inverse seesaw mechanism is not present. To achieve this, we forbid the $\epsilon$ coupling in front of the singlet mass term $\mathbf{1}_F\mathbf{1}_F$ in equation~\eqref{eq_superpotential_2} by global charge assignments. Based on charges in Table~\ref{table:global-charges}, the Yukawa part of the superpotential $W_{\text{Yuk}}^{\text{II}}$ is now
\begin{align}
    \begin{split}
	W_\mathrm{Yuk}^{\text{II}} &= m_{10} \mathbf{10}_F^2 + \kappa \mathbf{10}_F^2\cdot\mathbf{54}_H+ \mathbf{45}_F^2\cdot(Z_1\mathbf{210}_H+Z_2\mathbf{54}'_H)\\
	&\quad + y_1 \mathbf{16}_F^2 \cdot \mathbf{10}_H \\
	&\quad + Y_1 \mathbf{16}_F \cdot \mathbf{1}_F \cdot \overline{\mathbf{16}}_H + Y_2 \mathbf{16}_F \cdot \mathbf{10}_F \cdot \mathbf{16}_H + Y_3 \mathbf{16}_F \cdot \mathbf{45}_F \cdot \overline{\mathbf{16}}_H \;.
	\end{split}\label{eq_superpotential_model_II}
\end{align}

The terms which are still forbidden are $\mathbf{1}_F\mathbf{10}_F\mathbf{10}_H$ and $\mathbf{45}_F\mathbf{10}_F\mathbf{10}_H$, while the newly forbidden terms are $\mathbf{1}_F^2$ and $\mathbf{45}_F^2$, i.e.~$m_{45}=0$ and $\epsilon=0$. Furthermore, the term $\mathbf{45}_F\mathbf{1}_F\mathbf{45}_H$, which is dangerous for spoiling neutrino mass generation by strongly coupling the upper-left $3\times 3$ block in equation~\eqref{eq_neutrino_mass_matrix} to the decoupled states in $\mathbf{24}$ and $\mathbf{1}$ parts of the $\mathbf{45}_F$, is also not allowed by the charge assignments.

Since the mass term for the $\mathbf{45}_F$ is now forbidden by the charges, the $\mathbf{24}$ and $\mathbf{1}$ parts of $\mathbf{45}_F$ are now decoupled from the vector-like states by having the $Z_2$ coupling to the $\mathbf{54}_H'$ in
addition to the $Z_1$ coupling to $\mathbf{210}_H$. The VEV of $\mathbf{210}_H$ now has to be aligned in the $\mathbf{24}$ direction of $\mathbf{SU}(5)$, the same direction as the VEV of $\mathbf{54}'_H$.
Tuning in $Z_1$ and $Z_2$ then allows only the vector-like states to have $\mathrm{TeV}$ scale masses. Also, as will be seen later, $v_L=0$ in this case. With these considerations, the mass matrices in Eq.~\eqref{eq:ude_mass_matrix} are now replaced with
\begin{align}
    \mathbf{M}_u&=
        \begin{pmatrix}
         y_1 v_u & -Y_{3}\bar{v}_L & Y_3 \bar{v}_R \\
         -Y_3\bar{v}_L  & 0 & m_{45}'  \\
         Y_3 \bar{v}_R & -4 m_{45}'  & 0 \\
        \end{pmatrix},
    &\mathbf{M}_d&=
        \begin{pmatrix}
        y_1 v_d& 0 & Y_3 \bar{v}_R \\
        0 & 0 & m_{45}' \\
        Y_2 v_R & m_{10}'  & 0 \\
        \end{pmatrix},
    &\mathbf{M}_e&=
        \begin{pmatrix}
         y_1 v_d & 0 & Y_2 v_R \\
        0 & 0 & m_{10}''  \\
        Y_3 \bar{v}_R & 6 m_{45}' & 0 \\
        \end{pmatrix}.\label{eq:ude-II}
\end{align}
They are analogous to the previous result, except the some of the previously equal $\mathrm{TeV}$ scale parameters can now be different. Mixing between the SM particles and vector-like states remains present.

The neutrino mass matrix in Eq.~\eqref{eq_neutrino_mass_matrix} is replaced with one where $v_L=0$ and with no $\epsilon$ in the $(3,3)$ entry:
\begin{align}
    \mathbf{M}_\nu &=
    \begin{pmatrix}
     0 & y_1 v_u & Y_1 \bar{v}_L & 0 & Y_2 v_R \\
    y_1 v_u & 0 & Y_1 \bar{v}_R & 0 & 0 \\
    Y_1 \bar{v}_L & Y_1 \bar{v}_R & 0 & 0 & 0 \\
    0 & 0 & 0 & 0 & m_{10}''  \\
    Y_2 v_R & 0 & 0 & m_{10}''  & 0 \\
    \end{pmatrix} \;.
\end{align}
Above, we have used the definitions
\begin{align}
m_{10}'&:=m_{10}+2\kappa\, \langle\mathbf{54}\rangle,\label{eq:10p}\\
m_{10}''&:=m_{10}-3\kappa\, \langle\mathbf{54}\rangle,\label{eq:10pp}\\
m_{45}'&:=Z_1\,\langle \mathbf{210}_H\rangle_{24} - Z_2\,\langle\mathbf{54}'_H\rangle,
\end{align}
and we assume that $m_{10}$ and $\kappa$ are small enough that both $m_{10}'$ and $m_{10}''$ are in the $\mathrm{TeV}$ range. Similarly, we tune $Z_1$ and $Z_2$ so that $m_{45}'$ is in the $\mathrm{TeV}$ range as well, while the mass entries for the $\mathbf{24}$ and $\mathbf{1}$ parts of $\mathbf{45}_F$ have a different Clebsches and thus a linear combination, achieving decoupling from vector-like states.

Finally, for doublet-triplet splitting, the relevant part of the superpotential $W_{\text{DT}}$with terms allowed by the $\mathbb{Z}_4$ charges of Example II is
\begin{align}
W_{\text{DT}}&=m_{10}^H\,\mathbf{10}_H^2+\eta'\,\mathbf{10}_H^2\cdot\mathbf{54}_H^2 + \eta\,\mathbf{16}_H\cdot\mathbf{\overline{16}}_H\cdot\mathbf{45}_H+\lambda\,\mathbf{16}_H^2\cdot\mathbf{10}_H.
\end{align}
The term $\mathbf{\overline{16}}_H\cdot\mathbf{10}_H$ is forbidden by global charges, i.e.~$\bar{\lambda}=0$ in Eq.~\eqref{eq:mass-doublets}. Consequently, the induced EW breaking VEV in $\mathbf{16}_H$ is vanishing: $v_L=0$. The relevant induced EW symmetry breaking VEV $\bar{v}_L$ in $\mathbf{\overline{16}}_H$, however, is controlled by the $\lambda$ coupling, so the doublet-triplet splitting considerations of Appendix~\ref{A} and Eq.~\eqref{eq_linear_seesaw_2} and \eqref{eq_linear_seesaw_3} for linear seesaw apply.

Example II thus achieves everything we set out to do: neutrino mass is generated via linear seesaw, with no inverse seesaw contribution in this model. Furthermore, we achieve doublet-triplet splitting and the decoupling of the $\mathbf{24}$ and $\mathbf{1}$ parts under $\mathrm{SU}(5)$, while maintaining mixing of SM particles with vector-like states in the up, down and strange lepton sectors. Despite $v_L=0$, this mixing is still present both in the rows and columns of the mass matrices of Eq.~\eqref{eq:ude-II}, so the considerations of mixings of left-handed leptons and quarks with vector-like states of Section~\ref{section:constraints} still applies.

\subsection{Example III: inverse-seesaw only\label{Appendix:III}}

This model is merely an illustration how one can achieve a pure inverse-seesaw mechanism without any linear seesaw contribution. In this example, there is no mixing between the left-handed lepton fields of the SM and the vector-like family, therefore it is not a relevant example from the point of view of Section~\ref{section:constraints}.

In Example III of Table~\ref{table:global-charges}, the Higgs sector is extended by the representations $\mathbf{54}_H$, $\mathbf{45}_H$ and $\mathbf{210}_H$. As in Example I, the first two are required for doublet triplet splitting, while the $\mathbf{210}$ is used to decouple the $\mathbf{24}$ and $\mathbf{1}$ parts under $\mathrm{SU}(5)$ to be heavy in the $\mathbf{45}_F$. In contrast with Example I, global charge assignments forbid the term $\mathbf{16}_H^2\cdot\mathbf{10}_H$. As per Eq.~\eqref{eq_linear_seesaw_2} the induced VEV $\bar{v}_L$ is then not generated, giving no linear seesaw contribution to neutrino masses.

More concretely, the Yukawa part of the superpotential $W_{\text{Yuk}}^{\text{III}}$ is
\begin{align}
    \begin{split}
	W_\mathrm{Yuk}^{\text{III}} &= \epsilon \mathbf{1}_F^2+ m_{10} \mathbf{10}_F^2 + \kappa \mathbf{10}_F^2\cdot\mathbf{54}_H+ m_{45}\mathbf{45}_F^2+Z_1\mathbf{45}_F^2\cdot\mathbf{210}_H\\
	&\quad + y_1 \mathbf{16}_F^2 \cdot \mathbf{10}_H \\
	&\quad + Y_1 \mathbf{16}_F \cdot \mathbf{1}_F \cdot \overline{\mathbf{16}}_H + Y_3 \mathbf{16}_F \cdot \mathbf{45}_F \cdot \overline{\mathbf{16}}_H \;,
	\end{split}\label{eq_superpotential_model_III}
\end{align}
and the relevant part of the superpotential for doublet-triplet splitting $W_{\text{DT}}$ is
\begin{align}
W_{\text{DT}}&=m_{10}^H\,\mathbf{10}_H^2+\eta'\,\mathbf{10}_H^2\cdot\mathbf{54}_H^2 + \eta\,\mathbf{16}_H\cdot\mathbf{\overline{16}}_H\cdot\mathbf{45}+\bar{\lambda}\,\mathbf{\overline{16}}_H^2\cdot\mathbf{10}_H.
\end{align}
Compared to Example II in Appendix~\ref{Appendix:II}, the Yukawa sector now has the mass terms for $\mathbf{1}_F$ and $\mathbf{45}_F$ reintroduced, the $\mathbf{54}'$ representation is no longer there ($Z_2=0$) and also mixing between $\mathbf{16}_F$ and $\mathbf{10}_F$ is now forbidden, i.e.~$Y_2=0$. In $W_{\text{DT}}$, the $\lambda$ coupling is now forbidden, but $\bar{\lambda}$ is reintroduced. Consequently, based on the doublet mass matrix in Eq.~\eqref{eq:mass-doublets}, $\bar{v}_L=0$ and $v_L\neq 0$. The linear seesaw mechanism contribution thus vanishes, as seen from Eq.~\eqref{eq_linear_seesaw_2}, and only inverse seesaw with $\epsilon\neq 0$ remains.

The mass matrices in Eq.~\eqref{eq:ude_mass_matrix} are now replaced with
\begin{align}
    \mathbf{M}_u&=
        \begin{pmatrix}
          y_1 v_u & 0 & Y_3 \bar{v}_R \\
            0 & 0 & m_{45}'  \\
            Y_3 \bar{v}_R & m_{45}'  & 0 \\
        \end{pmatrix},
    &\mathbf{M}_d&=
        \begin{pmatrix}
        y_1 v_d & 0 & Y_3 \bar{v}_R \\
        0 & 0 & m_{45}'  \\
        0 & m_{10}' & 0 \\
        \end{pmatrix},
    &\mathbf{M}_e&=
        \begin{pmatrix}
        y_1 v_d & 0 & 0 \\
        0 & 0 & m_{10}'' \\
        Y_3 \bar{v}_R & m_{45}' & 0 \\
        \end{pmatrix},
\end{align}
while the neutrino mass matrix from Eq.~\eqref{eq_neutrino_mass_matrix} is now
\begin{align}
    \mathbf{M}_\nu &=
    \begin{pmatrix}
     0 & y_1 v_u & 0 & 0 & 0 \\
    y_1 v_u & 0 & Y_1 \bar{v}_R & 0 & 0 \\
    0 & Y_1 \bar{v}_R & \epsilon  & 0 & 0 \\
    0 & 0 & 0 & 0 & m_{10}''  \\
    0 & 0 & 0 & m_{10}''  & 0 \\
    \end{pmatrix} \;,
\end{align}
with definitions for $m_{10}'$ and $m_{10}''$ as in Eq.~\eqref{eq:10p} and \eqref{eq:10pp}, respectively, while $m_{45}'$ is now defined by
\begin{align}
m_{45}'&:=m_{45}+Z_1\,\langle\mathbf{210}_H\rangle_{1}.
\end{align}
The VEV of the $\mathbf{210}_H$ is now assumed to be in the singlet direction of $\mathrm{SU}(5)$, and tuning $m_{45}$ and $Z_1$ yields a $m_{45}'$ at the $\mathrm{TeV}$ scale, while the $\mathbf{24}$ and $\mathbf{1}$ parts in $\mathbf{45}_F$ remain heavy at the GUT scale. Doublet triplet splitting is achieved as well, and there is inverse but no linear seesaw. The $5\times 5$ neutrino matrix is now block diagonal, while the mixing between the SM particles and the new vector-like states in the down and charged lepton sectors is now compromised.


\begin{thebibliography}{99}
 \setlength{\itemsep}{0em}

\bibitem{Langacker:2008yv}
  P.~Langacker,
  Rev.\ Mod.\ Phys.\  {\bf 81} (2009) 1199
  doi:10.1103/RevModPhys.81.1199
  [arXiv:0801.1345 [hep-ph]].

\bibitem{King:2017cwv}
  S.~J.~D.~King, S.~F.~King and S.~Moretti,
  arXiv:1712.01279 [hep-ph].



\bibitem{Descotes-Genon:2013wba}
  S.~Descotes-Genon, J.~Matias and J.~Virto,
  Phys.\ Rev.\ D {\bf 88} (2013) 074002
  doi:10.1103/PhysRevD.88.074002
  [arXiv:1307.5683 [hep-ph]];
  W.~Altmannshofer and D.~M.~Straub,
  Eur.\ Phys.\ J.\ C {\bf 73} (2013) 2646
  doi:10.1140/epjc/s10052-013-2646-9
  [arXiv:1308.1501 [hep-ph]];
  D.~Ghosh, M.~Nardecchia and S.~A.~Renner,
  JHEP {\bf 1412} (2014) 131
  doi:10.1007/JHEP12(2014)131
  [arXiv:1408.4097 [hep-ph]].

\bibitem{Aaij:2014ora}
  R.~Aaij {\it et al.} [LHCb Collaboration],
  Phys.\ Rev.\ Lett.\  {\bf 113} (2014) 151601
  doi:10.1103/PhysRevLett.113.151601
  [arXiv:1406.6482 [hep-ex]].

  \bibitem{Bifani}
S.~Bifani for the LHCb Collaboration, {\it Search for new physics with $b \to s \ell^+ \ell^-$ decays at LHCb}, CERN Seminar, 18 April 2017,
{\tt https://cds.cern.ch/record/2260258}.


\bibitem{Hiller:2017bzc}
  G.~Hiller and I.~Nisandzic,
  arXiv:1704.05444 [hep-ph];
  L.S. Geng, B.~Grinstein, S.~Jager, J.~Martin Camalich, X.L. Ren and R.X. Shi,
  arXiv:1704.05446 [hep-ph];
  B.~Capdevila, A.~Crivellin, S.~Descotes-Genon, J.~Matias and J.~Virto,
  arXiv:1704.05340 [hep-ph];
  D.~Ghosh,
  arXiv:1704.06240 [hep-ph];
  D.~Bardhan, P.~Byakti and D.~Ghosh,
  arXiv:1705.09305 [hep-ph].



\bibitem{Glashow:2014iga}
  S.~L.~Glashow, D.~Guadagnoli and K.~Lane,
  Phys.\ Rev.\ Lett.\  {\bf 114} (2015) 091801
  doi:10.1103/PhysRevLett.114.091801
  [arXiv:1411.0565 [hep-ph]].


\bibitem{DAmico:2017mtc}
  G.~D'Amico, M.~Nardecchia, P.~Panci, F.~Sannino, A.~Strumia, R.~Torre and A.~Urbano,
  arXiv:1704.05438 [hep-ph].


  \bibitem{large}
 R.~Gauld, F.~Goertz and U.~Haisch,
  JHEP {\bf 1401} (2014) 069
  doi:10.1007/JHEP01(2014)069
  [arXiv:1310.1082 [hep-ph]];
  A.~J.~Buras and J.~Girrbach,
  JHEP {\bf 1312} (2013) 009
  doi:10.1007/JHEP12(2013)009
  [arXiv:1309.2466 [hep-ph]];
 A.~J.~Buras, F.~De Fazio and J.~Girrbach,
  JHEP {\bf 1402} (2014) 112
  doi:10.1007/JHEP02(2014)112
  [arXiv:1311.6729 [hep-ph]];
  W.~Altmannshofer, S.~Gori, M.~Pospelov and I.~Yavin,
  Phys.\ Rev.\ D {\bf 89} (2014) 095033
  doi:10.1103/PhysRevD.89.095033
  [arXiv:1403.1269 [hep-ph]];
  A.~Crivellin, G.~D'Ambrosio and J.~Heeck,
  Phys.\ Rev.\ Lett.\  {\bf 114} (2015) 151801
  doi:10.1103/PhysRevLett.114.151801
  [arXiv:1501.00993 [hep-ph]]
and  
  Phys.\ Rev.\ D {\bf 91} (2015) no.7,  075006
  doi:10.1103/PhysRevD.91.075006
  [arXiv:1503.03477 [hep-ph]];
  C.~Niehoff, P.~Stangl and D.~M.~Straub,
  Phys.\ Lett.\ B {\bf 747} (2015) 182
  doi:10.1016/j.physletb.2015.05.063
  [arXiv:1503.03865 [hep-ph]];
  A.~Celis, J.~Fuentes-Martin, M.~Jung and H.~Serodio,
  Phys.\ Rev.\ D {\bf 92} (2015) no.1,  015007
  doi:10.1103/PhysRevD.92.015007
  [arXiv:1505.03079 [hep-ph]];
 A.~Greljo, G.~Isidori and D.~Marzocca,
  JHEP {\bf 1507} (2015) 142
  doi:10.1007/JHEP07(2015)142
  [arXiv:1506.01705 [hep-ph]];
   W.~Altmannshofer and I.~Yavin,
  Phys.\ Rev.\ D {\bf 92} (2015) no.7,  075022
  doi:10.1103/PhysRevD.92.075022
  [arXiv:1508.07009 [hep-ph]];
 A.~Falkowski, M.~Nardecchia and R.~Ziegler,
  JHEP {\bf 1511} (2015) 173
  doi:10.1007/JHEP11(2015)173
  [arXiv:1509.01249 [hep-ph]];
  B.~Allanach, F.~S.~Queiroz, A.~Strumia and S.~Sun,
  Phys.\ Rev.\ D {\bf 93} (2016) no.5,  055045
  doi:10.1103/PhysRevD.93.055045
  [arXiv:1511.07447 [hep-ph]];
   C.~W.~Chiang, X.~G.~He and G.~Valencia,
  Phys.\ Rev.\ D {\bf 93} (2016) no.7,  074003
  doi:10.1103/PhysRevD.93.074003
  [arXiv:1601.07328 [hep-ph]];
  S.~M.~Boucenna, A.~Celis, J.~Fuentes-Martin, A.~Vicente and J.~Virto,
  Phys.\ Lett.\ B {\bf 760} (2016) 214
  doi:10.1016/j.physletb.2016.06.067
  [arXiv:1604.03088 [hep-ph]]
and
  JHEP {\bf 1612} (2016) 059
  doi:10.1007/JHEP12(2016)059
  [arXiv:1608.01349 [hep-ph]];
  P.~Ko, Y.~Omura, Y.~Shigekami and C.~Yu,
  arXiv:1702.08666 [hep-ph].
  K.~Ishiwata, Z.~Ligeti and M.~B.~Wise,
  JHEP {\bf 1510} (2015) 027
  doi:10.1007/JHEP10(2015)027
  [arXiv:1506.03484 [hep-ph]];
  D.~Aristizabal Sierra, F.~Staub and A.~Vicente,
  Phys.\ Rev.\ D {\bf 92} (2015) no.1,  015001
  doi:10.1103/PhysRevD.92.015001
  [arXiv:1503.06077 [hep-ph]];
  G.~Belanger, C.~Delaunay and S.~Westhoff,
  Phys.\ Rev.\ D {\bf 92} (2015) 055021
  doi:10.1103/PhysRevD.92.055021
  [arXiv:1507.06660 [hep-ph]];
  C.~Bobeth, A.~J.~Buras, A.~Celis and M.~Jung,
  JHEP {\bf 1704} (2017) 079
  doi:10.1007/JHEP04(2017)079
  [arXiv:1609.04783 [hep-ph]].
  A.~Crivellin, G.~D'Ambrosio and J.~Heeck,
  Phys.\ Rev.\ D {\bf 91} (2015) no.7,  075006
  doi:10.1103/PhysRevD.91.075006
  [arXiv:1503.03477 [hep-ph]].
  A.~Crivellin, J.~Fuentes-Martin, A.~Greljo and G.~Isidori,
  Phys.\ Lett.\ B {\bf 766} (2017) 77
  doi:10.1016/j.physletb.2016.12.057
  [arXiv:1611.02703 [hep-ph]].
  A.~Carmona and F.~Goertz,
  Phys.\ Rev.\ Lett.\  {\bf 116} (2016) no.25,  251801
  doi:10.1103/PhysRevLett.116.251801
  [arXiv:1510.07658 [hep-ph]].
  A.~Carmona and F.~Goertz,
  arXiv:1712.02536 [hep-ph].


 \bibitem{recent}
  C.~W.~Chiang, X.~G.~He, J.~Tandean and X.~B.~Yuan,
  arXiv:1706.02696 [hep-ph];
  Y.~Tang and Y.~L.~Wu,
  arXiv:1705.05643 [hep-ph];
  F.~Bishara, U.~Haisch and P.~F.~Monni,
  arXiv:1705.03465 [hep-ph];
  J.~Ellis, M.~Fairbairn and P.~Tunney,
  arXiv:1705.03447 [hep-ph];
  J.~F.~Kamenik, Y.~Soreq and J.~Zupan,
  arXiv:1704.06005 [hep-ph];
  F.~Sala and D.~M.~Straub,
  arXiv:1704.06188 [hep-ph];
  S.~Di Chiara, A.~Fowlie, S.~Fraser, C.~Marzo, L.~Marzola, M.~Raidal and C.~Spethmann,
  arXiv:1704.06200 [hep-ph];
  A.~K.~Alok, B.~Bhattacharya, A.~Datta, D.~Kumar, J.~Kumar and D.~London,
  arXiv:1704.07397 [hep-ph];
  R.~Alonso, P.~Cox, C.~Han and T.~T.~Yanagida,
  arXiv:1704.08158 [hep-ph];
  C.~Bonilla, T.~Modak, R.~Srivastava and J.~W.~F.~Valle,
  arXiv:1705.00915 [hep-ph];
  S.~F.~King,
  JHEP {\bf 1708} (2017) 019
  doi:10.1007/JHEP08(2017)019
  [arXiv:1706.06100 [hep-ph]].


\bibitem{Langacker:1988ur}
  P.~Langacker and D.~London,
  Phys.\ Rev.\ D {\bf 38} (1988) 886.
  doi:10.1103/PhysRevD.38.886


\bibitem{King:2017anf}
  S.~F.~King,
  JHEP {\bf 1708} (2017) 019
  doi:10.1007/JHEP08(2017)019
  [arXiv:1706.06100 [hep-ph]].

\bibitem{Malinsky:2005bi}
  M.~Malinsky, J.~C.~Romao and J.~W.~F.~Valle,
  Phys.\ Rev.\ Lett.\  {\bf 95} (2005) 161801
  doi:10.1103/PhysRevLett.95.161801
  [hep-ph/0506296].

\bibitem{Dev:2009aw}
  P.~S.~B.~Dev and R.~N.~Mohapatra,
  Phys.\ Rev.\ D {\bf 81} (2010) 013001
  doi:10.1103/PhysRevD.81.013001
  [arXiv:0910.3924 [hep-ph]].


\bibitem{Hisano:2015pma}
  J.~Hisano, Y.~Muramatsu, Y.~Omura and M.~Yamanaka,
  Phys.\ Lett.\ B {\bf 744} (2015) 395
  doi:10.1016/j.physletb.2015.04.020
  [arXiv:1503.06156 [hep-ph]].

\bibitem{typeI}
P.~Minkowski,
Phys.\ Lett.\ B {\bf 67} (1977) 421;
T. Yanagida, ``Horizontal symmetry and masses of neutrinos,'' In Proceedings of the
Workshop on the Baryon Number of the Universe and Unified Theories, Tsukuba, Japan,
13-14 Feb 1979;
M. Gell-Mann, P. Ramond and R. Slansky, ``Complex Spinors And Unified Theories,''
in Super- gravity, P. van Nieuwenhuizen and D.Z. Freedman (eds.), North Holland Publ. Co., 1979;
S. L. Glashow, NATO Adv. Study Inst. Ser. B Phys. 59 (1979) 687;
R.~N.~Mohapatra and G.~Senjanovi\'c,
Phys.\ Rev.\ Lett.\ {\bf 44} (1980) 912.


\bibitem{typeII}
M.~Magg and C.~Wetterich,
Phys.\ Lett.\ B {\bf 94} (1980) 61;
J.~Schechter and J.~W.~F.~Valle,
Phys.\ Rev.\ D {\bf 22} (1980) 2227;
G.~Lazarides, Q.~Shafi and C.~Wetterich,
Nucl.\ Phys.\ B {\bf 181} (1981) 287;
R.~N.~Mohapatra and G.~Senjanovi\'c,
Phys.\ Rev.\ D {\bf 23} (1981) 165.


\bibitem{typeIII}
R.~Foot, H.~Lew, X.~G.~He and G.~C.~Joshi,
Z.\ Phys.\ C {\bf 44} (1989) 441.


\bibitem{Capdevila:2017bsm}
  B.~Capdevila, A.~Crivellin, S.~Descotes-Genon, J.~Matias and J.~Virto,
  JHEP {\bf 1801} (2018) 093
  doi:10.1007/JHEP01(2018)093
  [arXiv:1704.05340 [hep-ph]].

\bibitem{Altmannshofer:2017yso}
  W.~Altmannshofer, P.~Stangl and D.~M.~Straub,
  Phys.\ Rev.\ D {\bf 96} (2017) no.5,  055008
  doi:10.1103/PhysRevD.96.055008
  [arXiv:1704.05435 [hep-ph]].

\bibitem{Bhattacharya:2016mcc}
  B.~Bhattacharya, A.~Datta, J.~P.~Gu\'{e}vin, D.~London and R.~Watanabe,
  JHEP {\bf 1701} (2017) 015
  doi:10.1007/JHEP01(2017)015
  [arXiv:1609.09078 [hep-ph]].

\bibitem{Amhis:2014hma}
  Y.~Amhis {\it et al.} [Heavy Flavor Averaging Group (HFAG)],
  arXiv:1412.7515 [hep-ex].

\bibitem{Cline:2017lvv}
  J.~M.~Cline, J.~M.~Cornell, D.~London and R.~Watanabe,
  Phys.\ Rev.\ D {\bf 95} (2017) no.9,  095015
  doi:10.1103/PhysRevD.95.095015
  [arXiv:1702.00395 [hep-ph]].


\bibitem{Osland:2017ema}
  P.~Osland, A.~A.~Pankov and A.~V.~Tsytrinov,
  Phys.\ Rev.\ D {\bf 96} (2017) no.5,  055040
  doi:10.1103/PhysRevD.96.055040
  [arXiv:1707.02717 [hep-ph]].

\bibitem{Aaboud:2017buh}
  M.~Aaboud {\it et al.} [ATLAS Collaboration],
  arXiv:1707.02424 [hep-ex].

\bibitem{sumrule}
S.~F.~King,
JHEP {\bf 0508} (2005) 105
[arXiv:hep-ph/0506297];
I.~Masina,
Phys.\ Lett.\  B {\bf 633} (2006) 134
[arXiv:hep-ph/0508031];
  S.~Antusch and S.~F.~King,
  Phys.\ Lett.\  B {\bf 631} (2005) 42
  [arXiv:hep-ph/0508044];
S.~Antusch, P.~Huber, S.~F.~King and T.~Schwetz,
JHEP {\bf 0704} (2007) 060
[arXiv:hep-ph/0702286];
  S.~Antusch, C.~Gross, V.~Maurer and C.~Sluka,
  Nucl.\ Phys.\ B {\bf 866} (2013) 255
  [arXiv:1205.1051 [hep-ph]].

\bibitem{Agashe:2014kda}
  K.~A.~Olive {\it et al.} [Particle Data Group],
  Chin.\ Phys.\ C {\bf 38} (2014) 090001.
  doi:10.1088/1674-1137/38/9/090001

\bibitem{Kitano:2002mt}
  R.~Kitano, M.~Koike and Y.~Okada,
  Phys.\ Rev.\ D {\bf 66} (2002) 096002
   Erratum: [Phys.\ Rev.\ D {\bf 76} (2007) 059902]
  doi:10.1103/PhysRevD.76.059902, 10.1103/PhysRevD.66.096002
  [hep-ph/0203110].

\bibitem{Blondel:2013ia}
  A.~Blondel {\it et al.},
  arXiv:1301.6113 [physics.ins-det].

\bibitem{Knoepfel:2013ouy}
  K.~Knoepfel {\it et al.} [mu2e Collaboration],
  arXiv:1307.1168 [physics.ins-det].

\bibitem{Antusch:2016ejd}
  S.~Antusch, E.~Cazzato and O.~Fischer,
  Int.\ J.\ Mod.\ Phys.\ A {\bf 32} (2017) no.14,  1750078
  doi:10.1142/S0217751X17500786
  [arXiv:1612.02728 [hep-ph]].

\bibitem{Deppisch:2015qwa}
  F.~F.~Deppisch, P.~S.~Bhupal Dev and A.~Pilaftsis,
  New J.\ Phys.\  {\bf 17} (2015) no.7,  075019
  doi:10.1088/1367-2630/17/7/075019
  [arXiv:1502.06541 [hep-ph]].
  
  \end{thebibliography}
\end{document}